\documentclass[12pt, draftclsnofoot, onecolumn]{IEEEtran}
\pdfoutput=1
\usepackage{amsmath}
\usepackage{amssymb,amsthm}
\usepackage{graphicx}
\usepackage{color}
\usepackage{bbm}
\usepackage[latin1]{inputenc}
\usepackage{verbatim}
\usepackage[update,prepend]{epstopdf}
\usepackage{mathrsfs}
\newtheorem{theorem}{Theorem}
\newtheorem{corollary}{Corollary}
\newtheorem{example}{Example}

\newcommand{\subsubsubsection}[1]{\textbf{#1: } }

\newcommand{\ind}[1]{\mathbf{1}\left(#1\right)}

\def\snr {\mbox{\scriptsize\sf SIR}}
\def\snr {\mbox{\scriptsize\sf SNR}}
\def\sinr {\mbox{\scriptsize\sf SINR}}
\newcommand{\PP}{\mathbb{P}}
\newcommand{\E}{\mathbb{E}}

\newcommand{\dd}{{\rm d}}
\newcommand{\ie}{{\em i.e., }}

\newcommand{\R}{\stsets{R}}

\def\E{\mathbb{E}}

\def\P{\mathbb{P}}
\def\pc{p_c}

\def\ie{{\em i.e.}}

\def\R{\mathbb{R}}

\def\L{\mathcal{L}}

\def\i{\mathbbm{1}}					
\def\d{\mathrm{d}}

\def\N{\sigma^2}

\def\sinr{\mathtt{SINR}}			
\def\snr{\mathtt{SNR}}

\def\rX{\mathbf{X}}

\def\x{\mathbf{x}}
\def\y{\mathbf{y}}
\def\tP{p}
\def\Th{\tau}
\def\T{\tau}							
\def\EM{\mu}
\def\lmeasure{\ell}
\def\user{\mathrm{u}}
\def\pc{\mathrm{p_c}}
\def\rUz{\mathbf{B}_0}
\newcommand\setnot[1]{\mathsf{#1}}
\newcommand{\Ir}{I_\mathrm{R}}
\newcommand{\IR}{I_\mathrm{R}}

\newcommand{\rh}{H}
\newcommand{\rg}{G}

\newtheorem{cor}[corollary]{Corollary}\vspace{-1in}

\newcommand{\includefig}[4]{\begin{figure}\centering\includegraphics[width=.5\textwidth,trim=#4,clip=true]{Figs/#1}\caption{#2}\label{#3}\end{figure}} 
 \newcommand{\Ball}{\mathcal{B}}
 
  \def\Cfunc{\zeta}
 \def\K{k}
 
 \newcommand{\prob}[1]{\PP\left[#1\right]}
 \newcommand{\ths}{\mathrm{th}}
 \allowdisplaybreaks
 
\begin{document}

\title{A Primer on Cellular Network Analysis Using Stochastic Geometry}
\author{Jeffrey G. Andrews, Abhishek K. Gupta, Harpreet S. Dhillon
\thanks{J. G. Andrews and A. K. Gupta are with the University of Texas at Austin.  H. S. Dhillon is with Virginia Tech.  The contact author is J. G. Andrews, jandrews@ece.utexas.edu. The technical content stems primarily from the papers \cite{AndBac11,NovDhi13,SinZhaJ2015,DhiGan12,JoSan12}.   Revision Date: \today. }}
\maketitle

\vspace{-0.5in}
\begin{abstract}
This tutorial is intended as an accessible but rigorous first reference for someone interested in learning how to model and analyze cellular network performance using stochastic geometry.   In particular, we focus on computing the signal-to-interference-plus-noise ratio (SINR) distribution, which can be characterized by the coverage probability (the SINR CCDF) or the outage probability (its CDF).  We model base stations (BSs) in the network as a realization of a homogeneous Poisson point process of density $\lambda$, and compute the SINR for three main cases: the downlink, uplink, and  finally the multi-tier downlink, which is characterized by having $\K$ tiers of BSs each with a unique density $\lambda_i$ and transmit power $\tP_i$.   These three baseline results have been extensively extended to many different scenarios, and we conclude with a brief summary of some of those extensions.
\end{abstract}

\section{Introduction}

The high speed global cellular communication network is one of humanity's most impressive and practically important technologies, and it continues to rapidly evolve and improve with each new generation.  However, until just a few years ago, mathematical performance analysis of these networks was not possible without resorting to extreme simplifications such as the Wyner model \cite{Wyn94} which considered only one or two interfering cells, or related models such as \cite{GilJac91} that lumped all this interference into a single random variable that was then empirically fit to some distribution, such as a lognormal.  Many other studies, e.g. \cite{AloGol99,Lin92} assume all the interferers were at the same distance away, which clearly is not the case in actual cellular networks.  

The alternative to such simplified models has been simply to exhaustively simulate the networks to average out the many sources of randomness, such as the BS and user locations and fading distributions, as well as the noise.  These simulations can be extremely time consuming and error-prone.  Although system-level simulations will continue to be indispensable for cellular network analysis and design, the need for a complementary analytical approach for the purposes of benchmarking and comparison has long been compelling.  Analysis can also illuminate key dependencies in the system, and provide guidance on what features and trends to more closely inspect.  An obvious analogy is the computation of bit error probability (BEP) in noisy (possibly fading) channels for different modulation formats: although this too can be simply simulated, it has been indispensable to have formulas which characterize the expected BEP.  Generally, these expressions are in the form of a Gaussian Q-function integral, a special function so common that it is often considered closed-form.   As we will see, functions almost this simple can describe the outage probability of the entire cellular network despite the presence of an infinite number of random variables (BS locations, fading values), under fairly reasonable assumptions. 

This tutorial is intended to provide a concise introduction to the tools and techniques for computing the distribution of signal-to-interference-plus noise ratio (SINR) in three key cases for cellular networks.  The key new aspect of the model is that all the BSs are located according to a Poisson point process (PPP), which intuitively means they are randomly scattered in the plane with independent locations.  We focus on the calculation of the \emph{coverage probability} which gives the complementary cumulative distribution function (CCDF) of SINR and is the complement of the outage probability. Either one gives the entire SINR distribution. In particular, this tutorial covers the following topics.
\begin{enumerate}
\item  An introduction to point processes, and particularly the PPP, including key mathematical definitions and computational tools that will be needed to derive the coverage probability.

 \item  {\em The cellular downlink.}   We first derive an expression for the coverage probability in a macrocell-only downlink cellular network where all the BSs are located according to a PPP and their signals experience independent and identically distributed (i.i.d.) Rayleigh fading in addition to standard power-law path loss.  The mobile user associates with the closest BS which is equivalently the one with the strongest average power. Three special cases of increasing mathematical simplicity are given.
 
 \item {\em The cellular uplink.} The above approach is extended to an uplink cellular system model including transmit power control at the mobile user (i.e. handset).  This problem is more difficult than the downlink due to the coupling between the handset point process and the BS point process when it is assumed (realistically) that only a single handset can be active per cell (in a given time/frequency resource). Under some reasonable approximations, we are able to accurately characterize coverage probability in this case as well.
 
 \item {\em The HetNet downlink.}  We extend the downlink results to a heterogeneous cellular network (HCN, or ``HetNet'') with $\K$ tiers, where each tier is distinguished by a unique transmit power $\tP_i$, density $\lambda_i$, and SINR threshold $\Th_i$.  Such a model is quite general and can describe any overlaid collection of macro, micro, pico, and femtocells, which will become increasingly relevant in the future \cite{And5G}.   Despite this significant additional complexity in the model, we are able to compute its coverage probability in a fairly simple form. This result can be viewed as a generalization of the main result on coverage for macrocells only.
 
 \end{enumerate}

These three baseline models and their corresponding SINR derivations provide a broad framework to build upon in terms of developing tractable analytical models for cellular networks.  And indeed, these results have been extended in literally hundreds of ways to date.  We conclude the paper by discussing some of them.

\section{Key Background on Stochastic Geometry} \label{sec:background}
In this section, we concisely introduce a few basic concepts from stochastic geometry and some key results for Poisson point processes, with simple illustrative examples.  For a more thorough exposition, the interested reader is advised to consider \cite{HaenggiBook} or \cite{BacNOW}.

{\em Notation:} Throughout this tutorial, we denote the random variables by upper-case letters and their realizations or other deterministic quantities by lower-case letters. We use bold font to denote vectors and normal font to denote scalar quantities. Therefore, a random vector (e.g., a random location) will be denoted by an upper case letter in the bold font. Similarly, lower case letters in bold font will denote deterministic vectors. For instance, $X$ and $\rX$ denote one-dimensional (scalar) random variable and random vector (containing more than one element), respectively. Similarly, $x$ and $\x$ denote scalar and vector of deterministic values, respectively. The upper case letters in the san-serif font, such as $\setnot{A}$ and $\setnot{L}$, will be used to denote sets.

\subsection{Point Process Essentials}
Put simply, a point process (PP) $\Phi = \{\rX_i,i\in \mathbb{N}\}$ is a {\em random} collection of points residing in a measure space, which for cellular networks is the Euclidean space $\mathbb{R}^d$. One way to interpret $\Phi$ is in terms of the so called {\em random set formalism}, where $\Phi = \{\rX_i\} \subset \mathbb{R}^d$ is a {\em countable} random set with each element $\rX_i$ being a random variable. An equivalent and usually more convenient interpretation is in terms of the {\em random counting measure}, where the idea is to simply count the number of points falling in any set $\setnot{A} \subset \mathbb{R}^2$. It is mathematically defined as
\begin{align}
\Psi(\setnot{A})=\sum_{\rX_i\in \Phi} \i(\rX_i\in \setnot{A}).
\end{align}
Note that $\Psi(\setnot{A})$ is a random variable whose distribution depends upon $\Phi$. Clearly if $\Psi(\setnot{A})$ is exhaustively considered for {\em all} possible sets $\setnot{A}$, it can completely describe the PP $\Phi$. For a more rigorous introduction to PPs, please refer to \cite{HaenggiBook} or \cite{BacNOW}. Before proceeding further, let's look at a simple example of a PP and random counting measure below.


\begin{example}\label{ex:1}
Consider a point process $\Phi_\mathrm{S}$ in $\R^2$ which admits just two possible realizations: (i) $\Phi_\mathrm{S}=\phi_1$ with probability $\frac14$, which consists of two points at $\x_1=(1,0)$ and $\x_2=(0,1)$ and (ii) $\Phi_\mathrm{S}=\phi_2$ with probability $\frac34$ which consists of three points at  $\x_1=(0, 0)$, $\x_2=(1,1)$ and $\x_3=(2,2)$ (See Fig. \ref{fig:example1}). This point process can be equivalently characterized in terms of the random counting measure $\Psi(\cdot)$, which for any set $A$ takes two possible values: 
\begin{align*}
\Psi(\setnot{A}) = \psi_1(\setnot{A})&=\mathbbm{1}((1,0)\in \setnot{A})+\mathbbm{1}((0,1)\in \setnot{A}), \\
\Psi(\setnot{A}) = \psi_2(\setnot{A})&=\mathbbm{1}((0,0)\in \setnot{A})+\mathbbm{1}((1,1)\in \setnot{A}) +\mathbbm{1}((2,2)\in \setnot{A}) .
\end{align*} 
Clearly, $\Psi(\setnot{A}) = \psi_1(\setnot{A})$ with probability $\frac{1}{4}$, and $\Psi(\setnot{A}) = \psi_2(\setnot{A})$ with probability $\frac{3}{4}$. Now consider the set $\setnot{A}=\Ball((1,1),1.1)$ which is a ball of radius $r=1.1$ around the location $(1,1)$, which happens to be the location of a point in $\psi_2$. Then $\psi_1(\setnot{\setnot{A}})=2$ and $\psi_2(\setnot{A})=1$.
\end{example}

\includefig{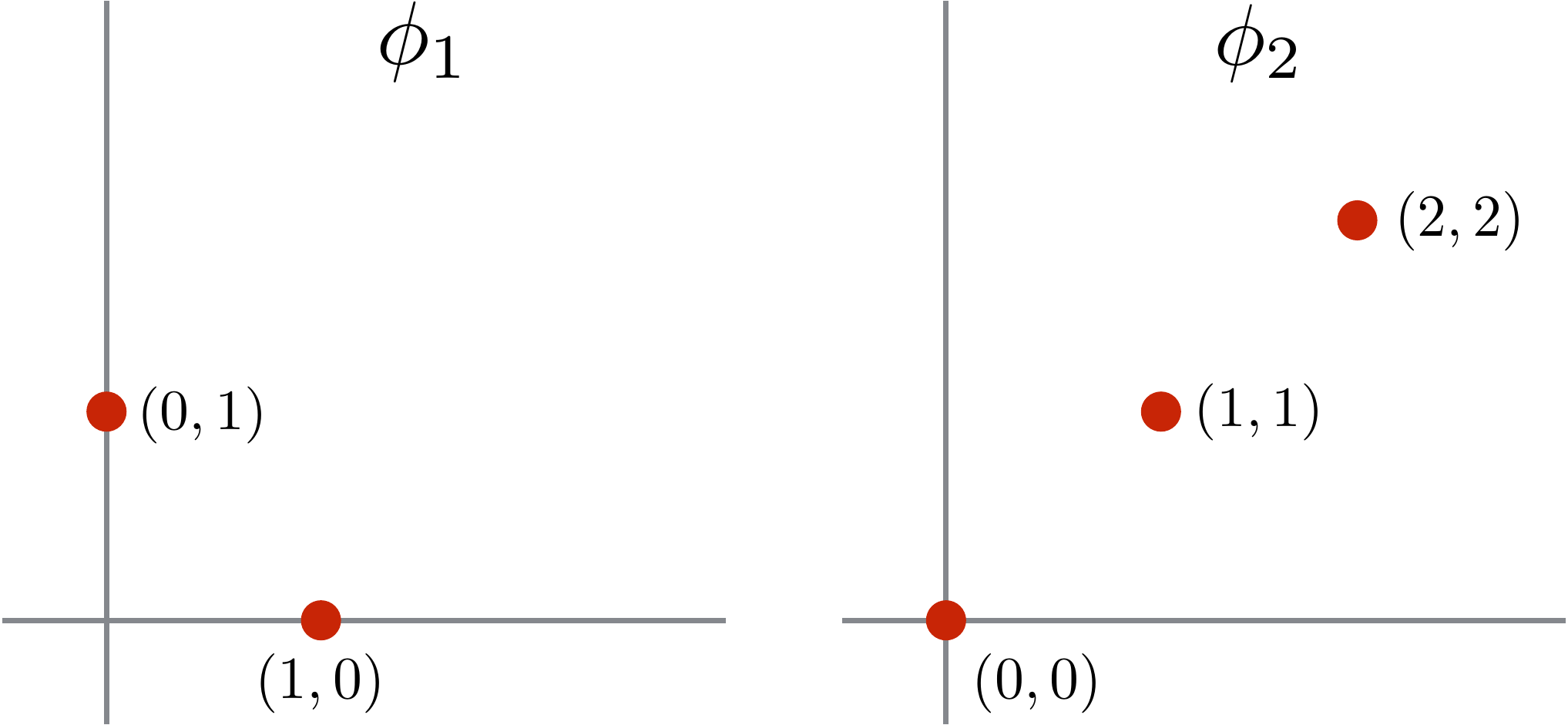}{Example 1: Two realizations $\phi_1$ (left) and $\phi_2$ (right) of the point process $\Phi_\mathrm{S}$.}{fig:example1}{0 0 0 0 }

A \emph{marked point process} is a point process where a random variable (known as a {\em mark})  $Q_i$ is associated with each point $\rX_i$. This mark can be a independent random variable with some distribution or it can be a function of the PP as seen from the point $\rX_i$ (e.g., a {\em feature} of a PP measured at $\rX_i$). For example, let  PP $\Phi=\{\rX_i,i\in \mathbb{N}\}$ denote the locations of the BSs. We can now assign a independent random transmit power $P_i\sim \exp(1)$ to each point. The combined PP $\Phi_\mathrm{M}=\{\rX_i,P_i\}$ is then a marked point process. 

Now, we discuss some of the important statistical measures of a point process.

\subsubsection{\textbf{Expectation measure}}
 The \emph{expectation measure} of a point process is a function which maps a set $\setnot{A}$ to the mean number of points in it. It is defined as the mean of $\Psi(\cdot)$:
\begin{align}
\EM(\setnot{A})=\mathbb{E}\left[\Psi(\setnot{A})\right].
\end{align}

For the PP $\Phi_\mathrm{S}$ in Example \ref{ex:1}, the expectation measure can be computed as
\begin{align*}
\EM(\setnot{A})=\frac14 \psi_1(\setnot{A})+\frac34\psi_2(\setnot{A}).
\end{align*}

\subsubsection{\textbf{Probability generating functional (PGFL)}}
Let $f:\mathbb{R}^d\rightarrow \mathbb{R}^+$ be a function. Then the PGFL of the PP with respect to this function is defined as the mean of the product of the function's values at each point of the PP, {\em i.e.}
\begin{align}
\mathcal{P}_{\Phi}(f)=\mathbb{E}\left[{\prod_{\rX_i\in \Phi}f(\rX_i)}\right].
\end{align}
\begin{example} 
Let $f(\x)=\|\x\|^2$ for a given point $\x\in\R^2$. For the PP $\Phi_\mathrm{S}$ in Example \ref{ex:1}, the PGFL of $f$ can be computed as
\begin{align*}
\mathcal{P}_{\Phi}(f)=\frac14 \cdot 1\cdot 1+\frac34\cdot 0\cdot2\cdot8=0.25.
\end{align*}
\end{example}

The PGFL can also be used to compute the Laplace transform of a random variable $F$ if it can be written in following form:
\begin{align}
F=\sum_{\rX_i\in \Phi}g(\rX_i)\label{eq:rform}.
\end{align}
In this case, its  Laplace transform  can be computed as 
\begin{align}
\mathcal{L}_{F}(s)=\mathbb{E}\left[e^{-sF}\right]=&\mathbb{E}\left[\exp\left({-\sum_{\rX_i\in \Phi}sg(\rX_i)}\right)\right]=\mathbb{E}\left[{\prod_{\rX_i\in \Phi}e^{-sg(\rX_i)}}\right].
\end{align}

The PGFL is particularly important in many wireless applications, where the Laplace transform of interference is usually an intermediate step in the characterization of SINR. The following example demonstrates how PGFL can be used to derive the Laplace transform of interference.


\begin{example}\label{ex:interference}
BSs are randomly deployed in an infinite 2D space $\R^2$ and the PP $\Phi_\mathrm{BS}=\{\rX_i,i\in \mathbb{N}\}$ denotes their locations.  The sum interference at any point $\y$ is a random variable which can be written as the summation of signals from each BS attenuated according to the standard power-law path loss model, meaning that the received power attenuates with distance $r = \|\x-\y\|$ as $r^{-\alpha}$, thus
\begin{align*}
I&=\sum_{\rX_i\in \Phi_\mathrm{BS} }\frac{P_i}{\|\rX_i-\y\|^\alpha}
\end{align*}
where $\alpha$ is the path loss exponent. Assuming $P_i = \tP$, the form is the same as \eqref{eq:rform}, hence the Laplace transform of the interference is given by the PGFL of the PP $\Phi_\mathrm{BS}$ with respect to the function $f(\x)=\tP \|\x-\y\|^{-\alpha}$. 
\end{example}

\subsubsection{\textbf{Palm distribution}}
The Palm distribution of a PP at a given location is its conditional distribution conditioned on the presence of a point at that location. Therefore, the Palm distribution represents how the PP would look when viewed from one of its atoms (points). It is useful in studying the properties of a PP as observed from one of its points (such as the distance of a point of a PP to its nearest point, or the average number of points in a ball of radius $r$ with its center at a point of PP). Since such properties can be interpreted as marks of the points where they are observed, they can be formally studied using the Palm distribution of the corresponding marks. 
Mathematically, the Palm distribution of the mark $Q$ for the stationary marked PP $\Phi=\{\rX_i,Q_i\}$ is
\begin{align}
\nu_Q(\setnot{L})&=\mathbb{P}^0\left[Q_0\in \setnot{L} \right]
=\frac{\E\left[\sum\limits_{\rX_i\in \Phi}\ind{\rX_i\in \setnot{A}}\ind{Q_i\in \setnot{L}}\right]}{\E\left[\sum\limits_{\rX_i\in \Phi}\ind{\rX_i\in \setnot{A}}\right]}
\end{align}
where $\setnot{L}$ is an event from sample space of marks. Here $\mathbb{P}^0\left[Q_0\in \cdot \right]$ denotes the  distribution of mark at origin conditioned on the fact that there is a point at the  origin.


\subsection{Poisson Point Processes}
A (general) Poisson point process (PPP) is a point process with expectation measure $\EM(\cdot)$ if
\begin{enumerate}
\item $\Psi(\setnot{A})$ is Poisson distributed with mean $\EM(\setnot{A})$ for every set $\setnot{A}$.
\item For any $m$ disjoint  sets $\setnot{A}_1,\cdots , \setnot{A}_m$, the random variables $\Psi(\setnot{A}_1),\cdots , \Psi(\setnot{A}_m)$ are independent.
\end{enumerate}

We will be most interested in a \emph{homogeneous} Poisson point process, which is a  PPP with uniform intensity $\lambda$
such that 
\begin{align}
\EM(\setnot{A})=\lambda \lmeasure(\setnot{A})
\end{align}
where $\lmeasure(\setnot{A})$ is a Lebesgue measure ({\em i.e.} size) of $\setnot{A}$. An important property of a homogeneous PPP is that conditioned on the number of points in $\setnot{A}$ (which is Poisson distributed with mean $\lambda \lmeasure(\setnot{A})$), all the points are independently and uniformly distributed in $\setnot{A}$.

\begin{example}
 \label{ex:simplePPP}
Consider a PPP in $\R^2$ with intensity $\lambda$ having units of points/area. If $\setnot{A}$ is a set denoting a circle of radius $r$, we would have $\lmeasure(\setnot{A})=\pi r^2$ and $\EM(\setnot{A})=\lambda \pi r^2$. The probability that there are $n$ points in $\setnot{A}$ is given by
\begin{align}
\mathbb{P}\left[\Psi(\setnot{A})=n\right]&=\exp(-\lambda \pi r^2)\frac{(\lambda \pi r ^2)^n}{n!}.
\end{align}
\end{example}
 
 We now list some of the important properties and statistical measures for the homogeneous PPP.
 
\subsubsection{\textbf{Campbell's theorem}}:
Campbell's theorem can be used to compute the expectation of a random variable  $F$ of the form \eqref{eq:rform} and thus provides a key tool to convert a sum into a (hopefully computable) integral. It states that
\begin{align}
\E[F]=\mathbb{E}\left[\sum_{\rX_i\in\Phi} f(\rX_i)\right]&=\int_{\R^d} \lambda f(\mathbf{x}) \dd \mathbf{x}.
\end{align}

In the following example, we will see how Campbell's theorem can be used to compute the mean interference in a cellular system.
\begin{example}\label{ex:meaninterference}
Goal: find the mean interference at origin from the BSs located in the ring $\setnot{A}:{a\le\|\x\|<b}$.  Assuming a homogenous PPP($\lambda$), power-law path loss, and constant transmit power $\tP$, the interference can be written as 
\begin{align}
I_\setnot{A}&=\sum_{\rX_i\in\Phi,a\le \|\rX_i\|<b}\frac{\tP}{\|\rX_i\|^\alpha}
=\sum_{\rX_i\in\Phi}\frac{\tP}{\|\rX_i\|^\alpha}\mathbbm{1}(a\le \|\rX_i\|<b).
\end{align}
Using Campbell's theorem, we get the following expression:
\begin{align}
\E\left[I_\setnot{A}\right]&=\E\left[\sum_{\rX_i\in\Phi}\frac{\tP}{\|\rX_i\|^\alpha}\mathbbm{1}\left(a\le \|\rX_i\|<b\right)\right]
=\lambda\int_{\R^2}\frac{\tP}{\|\mathbf{x}\|^\alpha}\mathbbm{1}(a\le \|\mathbf{x}\|<b)\dd \mathbf{x}.
\end{align}
Converting to polar coordinates, the mean interference can be evaluated as
\begin{align}
\E\left[I_\setnot{A}\right]&=\lambda\int_0^\infty\int_0^{2\pi}\frac{\tP}{r^\alpha}\mathbbm{1}(a\le r<b)\dd \theta r\dd r \\
& =2\pi\lambda\int_a^b\frac{\tP}{r^\alpha} r \dd r\\
&= \frac{2\pi\lambda \tP}{2-\alpha} \left( b^{2-\alpha} - a^{2-\alpha} \right) = \frac{2\pi\lambda \tP}{\alpha-2}\left(\frac{1}{a^{\alpha-2}}-\frac{1}{b^{\alpha-2}}\right).
\end{align}
\end{example}

A couple observations can be made from Example \ref{ex:meaninterference}.  Consider first $b \to \infty$, that is interference comes from the entire 2D plane outside of an ``exclusion zone'' of radius $a$.  In this case, we can observe that the interference is finite iff $\alpha > 2$.  This is intuitive: the average number of interferers grows quadratically with $b$ per Example \ref{ex:simplePPP}, so the interference each one contributes needs to reduce by more than an inverse square law if the sum interference is to remain finite.  From a practical point of view, it means that free space propagation ($\alpha = 2$) is not quite sufficient in a PPP network that extends infinitely in the plane.  Similarly, for any $a > 0$ (and $\alpha > 2$), the mean interference remains finite.  This simple example illustrates why virtually all stochastic geometry results require $\alpha > 2$ and $a > 0$ (or a modified path loss model like $(r+1)^{-\alpha}$ that avoids the singularity at $r=0$). 

\subsubsection{\textbf{PGFL}}
 The PGFL of a homogeneous Poisson point process is given as
 \begin{align}
 \mathcal{P}_\Phi(f)=\mathbb{E}\left[ {\prod_{\rX_i\in \Phi} f(\rX_i)}\right] = \exp\left(-\lambda \int_{\mathbb{R}^d} (1-f(\mathbf{x})) \dd \mathbf{x}\right).
\end{align}
The PGFL is  useful for converting an expectation of a product of the points in the PPP into a (hopefully computable) integral. As discussed earlier, the Laplace transform of a random variable $F$ (defined in \eqref{eq:rform}) can be evaluated using PGFL as
\begin{align}
 \mathbb{E}\left[e^{-sF}\right]=\mathbb{E}\left[ \exp\left({-\sum_{\rX_i\in \Phi} sg(\rX_i)}\right)\right] = \mathcal{P}_\Phi\left(e^{-sg}\right)=\exp\left(-\lambda \int_{\mathbb{R}^2} \left(1-e^{-sg(\x)}\right) \dd \x\right).
\end{align}

\begin{example}
Consider the PPP $\Phi$ from Example \ref{ex:meaninterference} and assume $\alpha=4$. The Laplace transform of the sum interference at the origin can be computed as
\begin{align}
 \mathbb{E}\left[e^{-sI}\right]=\exp\left(-2\pi \lambda \int_{0}^\infty \left(1-e^{-s\tP{x}^{-4}}\right) x\dd x\right)=\exp\left(-\pi\lambda\sqrt{\pi s\tP}\right).
\end{align}
\end{example}


\subsubsection{\textbf{Slivnyak's theorem}}
Slivnyak's theorem states that for a Poisson point process $\Phi$, because of the independence between all of the points, conditioning on a point at $\x$ does not change the distribution of the rest of the process. Mathematically, this can be thought of as removing an infinitesimally small area corresponding to a ball $\Ball(\x, \epsilon$) for $\epsilon \to 0$, since the distributions of points in all nonoverlapping regions are independent for a PPP.  This means that any property seen from a point $\x$ is the same whether or not we condition on having a point at $\x$ in $\Phi$. 

Slivnyak's theorem is quite simple, but it is important because it allows us to add a node to the PPP at any location we like, such as the origin or at a fixed distance from the origin, without changing its statistical properties.  In the context of a cellular downlink network, it allows us to treat the interference as coming from a PPP despite removing the serving BS from the PPP. We will discuss this in detail in Section \ref{sec:DL}.

\begin{example}
Consider a PPP in $\R^2$ with intensity $\lambda$ having units of points/Area. The goal is to compute the probability that a randomly chosen point is farther than $r$ from its nearest neighbor. Note that the above probability is the Palm probability as it is the view of PPP from an arbitrary chosen point. Using the notation for Palm,  the desired probability can be written as
\begin{align}
p&=\P^0\left[\min_{i\ne 0}(\|\x_i-\x_0\|)>r\right]
\end{align}
where the superscript $0$ denotes that it is conditioned on the fact that there is a point at $0$. Note that due to stationarity of homogeneous PPP, the position of the arbitrary chosen point will not change the probability. Now,  from Slivnyak theorem, $p$ will be equal to
\begin{align}
p&=\P\left[\min_{i}(\|\x_i\|)>r\right]
\end{align}
which is nothing but the void probability of the PPP and is given as $\exp\left(-\lambda\pi r^2\right)$.
\end{example}


\subsubsection{\textbf{Properties of a PPP}}
We now list a few other important properties of the PPP. 
\begin{enumerate}
\item [(i)] \textbf{Independent thinning} of a PPP results in a different PPP.   For example, if we independently assign random binary $\{0,1\}$ marks with $\PP[Q_i = 1] = q$ to each point in a PPP and collect all the points which are marked as $1$, this new PP will also be PPP now with intensity $q \lambda$.
\item [(ii)] \textbf{Superposition} of independent PPPs results in a PPP. Thus if we combine $m$ independent homogeneous PPPs characterized by intensities $\lambda_i, ~ i = 1, 2, ..., m$ to form a new PP, this new PP will also be a PPP, now with intensity $\sum_{i=1}^m \lambda_i$.
\item [(iii)] \textbf{Displacement} of a PPP results in a different PPP.  This means that if we displace each point of a PPP by some random law, for example by adding independent and identically distributed (iid) 2D Gaussian random variables to each point, the PP consisting of these new random points will also be PPP.
\end{enumerate}

\section{Downlink Analysis}
\label{sec:DL}

We now apply these powerful tools to the analysis of a downlink cellular network \cite{AndBac11}.  

\subsection{Downlink Model and Metrics}\label{sec:DLSysMod}
The key aspects of the downlink system model are as follows:
\begin{itemize}
\item BSs are located according to a homogeneous PPP $\Phi$ of intensity $\lambda$ in the Euclidean plane. 
\item Mobile users are independently located according to some other stationary point process $\Phi_\user$. 
\item The BSs transmit constantly with fixed power $\tP$ to a single desired mobile user on any particular time-frequency resource, i.e. orthogonal multiple access within a cell.  Therefore, the mobile user sees interference from all other BSs in the network but not from its own desired BS.
\item Signals attenuate with distance according to the standard power-law path loss propagation model with path loss exponent $\alpha > 2$.  Specifically, we assume that the average received power at distance $r$ is $\tP_{\rm rx}(r) = \tP r^{-\alpha}$.  Necessary propagation constants -- notably the path loss experienced at the reference distance $r=1$ -- are excluded for simplicity, but can easily be incorporated into either the transmit power $\tP$ or the additive noise power $\sigma^2$.
\item The $\snr = \frac{\tP}{\sigma^2}$ is defined to be the average received $\snr$ at a distance of $r = 1$, which is our way of handling the aforementioned missing constant.
\item Random channel effects are incorporated by a multiplicative random value $\rh$ for the desired signal and $\rg_i$ for interferer $i$ (see~\eqref{eq:SINR_DL} and \eqref{eq:Ir_DL}).  For simplicity we assume these all correspond to Rayleigh fading with mean 1, so $\rh$ and $\{\rg_i\}$ all are iid and follow an exponential distribution with mean $1$.  We note that more general fading/shadowing distributions can be accommodated as in \cite{AndBac11} and many subsequent papers, but with a loss of tractability and without much change to the results and system design insights. 
\item All analysis is for a typical mobile user at the origin.  This user connects to the BS that provides the highest average $\snr$ which is equivalently the one providing the highest average $\sinr$, as well as being the closest BS.  Note that a different BS could provide higher instantaneous $\sinr$ depending on the fading values. The instantaneous $\sinr$ based association will be discussed in the context of heterogeneous cellular networks in Section~\ref{sec:HCN}.
\end{itemize}

The goal of this section is to carefully derive the probability of coverage in a downlink cellular network. 
 The coverage probability is defined as
\begin{equation}
\pc(\Th,\lambda,\alpha) \triangleq \PP[\sinr > \Th],
\label{eq:CovDefn}
\end{equation}
which is  exactly the CCDF of SINR over the entire network, since the CDF gives $\PP[\sinr \leq \Th]$. It can be thought of equivalently as:
\begin{enumerate}
\item the probability that a randomly chosen user can achieve a target SINR $\Th$, 
\item the average fraction of users who at any time achieve SINR $\Th$, or 
\item the average fraction of the network area that is in ``coverage'' at any time. 
\end{enumerate}
 
 The SINR of the mobile user at a random distance $R$ from its associated BS can be expressed as
\begin{equation}
\sinr = \frac{\rh\tP R^{-\alpha}}{\sigma^2 + \IR},
\label{eq:SINR_DL}
\end{equation}
where
\begin{equation}
\IR= \sum_{\rX_i\in\Phi \setminus \{ \mathbf{B_o}\}} \rg_i \tP \|\rX_i\|^{-\alpha}
\label{eq:Ir_DL}
\end{equation}
is the cumulative interference from all the other BSs, excluding the tagged BS for the mobile user at the origin $\mathbf{o}$ which is denoted by $\mathbf{B_o}$.  Interfering BS $i$ is located at  $\rX_i$.

 \subsection{Distance to the Nearest Base Station}
\label{subsec:NN}
An important quantity is the random distance $R$ separating a typical user from its tagged BS.  Since each user communicates with the closest BS, no other BS can be closer than $R$, so all interfering BSs must be farther than $R$.  The probability density function (PDF) of $R$ can be derived using the simple fact that the null probability of a 2-D Poisson process in an area $a$ is $\exp(-\lambda a)$.
\begin{align}
\PP[R > r] & =  \PP[{\rm No ~ BS ~ closer ~ than ~ r}]=e^{-\lambda \pi r^2}.
\end{align}
Therefore, the CDF is $\PP[R \leq r]  =  F_R(r)  = 1 - e^{-\lambda \pi r^2}$ and the pdf can be found as
\begin{align}
f_R(r) &= \frac{\dd F_R(r)}{\dd r}= 2\pi \lambda r e^{-\lambda \pi r^2}.
\end{align}

This Rayleigh distribution (in $r$) thus describes the nearest neighbor distance in a PPP, a well known classical result \cite{BacNOW}.

\subsection{Interference Characterization} \label{sec:Int_singletier}
The interference $\IR$ is a standard $M/M$ shot noise \cite{BacNOW,LowTei90} created by a Poisson point process of intensity $\lambda$ outside a disc centered at the origin $\mathbf{o}$ and of radius $R$, for which some useful results are known. To characterize the interference, we compute  the Laplace transform of random variable $\Ir$ at $s$ conditioned on the random distance $R=r$ to the closest BS from the origin, which we denote as 
${\cal L}_{\Ir}(s)$. The Laplace transform definition yields
\begin{align}
{\cal L}_{\Ir}(s) &= \E\left[e^{-s\IR}\right] = \E_{\Phi, \{\rg_i\}}\left[\exp\left(-s\sum_{\rX_i\in \Phi\setminus\{ \mathbf{B_o}\}} \rg_i \tP \|\rX_i\|^{-\alpha}\right)\right] \nonumber\\
&= \E_{\Phi, \{\rg_i\}}\left[\prod_{\rX_i\in \Phi\setminus\{ \mathbf{B_o}\}} \exp(-s \rg_i \tP \|\rX_i\|^{-\alpha})\right].
\end{align}
Now using the independence of the $\rg_i$'s, we can move the expectation with respect to $\rg_i$ inside the multiplication
\begin{align}
{\cal L}_{\Ir}(s) &=\E_{\Phi}\left[\prod_{\rX_i\in \Phi\setminus \{\mathbf{B_o}\}} \E_{\rg} [\exp(-s \rg\tP \|\rX_i\|^{-\alpha})] \right] .
\end{align}
Using the PGFL of PPP with respect to the function $f(\x)=\E_{\rg} [\exp(-s \rg \tP \|\x\|^{-\alpha})]$, we get
\begin{align}
{\cal L}_{\Ir}(s)&=  \exp\left(-\lambda \int_{\R^2\setminus\Ball(\mathbf{0},r)} \left(1 - \E_{\rg}[\exp(-s \rg\tP \|\x\|^{-\alpha})]  \right)   \dd \x \right).\end{align}
Employing a transformation to polar coordinates $\x=(x,\theta)$, we get
\begin{align}
{\cal L}_{\Ir}(s)&=  \exp\left(-2\pi\lambda \int_r^{\infty} \left(1 - \E_{\rg}[\exp(-s \rg \tP x^{-\alpha})]  \right) x  \dd x \right).
\label{eq:laplace}
\end{align}
The integration range excludes a ball centered at 0 and radius $r$ since the closest interferer has to be farther than the desired BS, which is at distance $r$.  Since $\rg_i \sim \exp(1)$, the moment generating function of an exponential random variable gives
\begin{align}
{\cal L}_{\Ir}(s) 
&=  \exp\left(-2\pi\lambda \int_r^{\infty} \left(1 - \frac{1}{1+s\tP x^{-\alpha}}\right) x  \dd x \right)\\
&=  \exp\left(-2\pi\lambda \int_r^{\infty} \left(\frac{1}{1+(s\tP)^{-1} x^{\alpha}}\right) x  \dd x \right)\label{eq:laplaceexp}.
\end{align}

Armed now with an expression for the Laplace transform of the interference, we proceed to the main result.

\subsection{Coverage Probability}
Conditioning on the nearest BS being at a distance $r$ from the typical user, the probability of coverage relative to an SINR threshold $\Th$ can be written as
\begin{align*}
\pc(\Th,\lambda,\alpha) &= \E_R\left[ \vphantom{\frac33} \PP[\sinr >\Th\ |\  R=r] \right]  = \int_{r>0} \PP[\sinr > \Th\ |\  r ]f_R(r)\dd r 
\end{align*}
Using the distribution $f_R(r)$ derived in Subsection \ref{subsec:NN}, we get
\begin{align*}
\pc(\Th,\lambda,\alpha) &=\int_{r>0} \PP\left[\frac{\rh\tP R^{-\alpha}}{\sigma^2+\IR} > \Th\ \Big| \ R=  r\right] e^{-\pi\lambda r^2} 2 \pi \lambda r \dd r \nonumber \\
& =  2\pi \lambda\int_{r>0} e^{-\pi \lambda r^2} \PP[\rh>\Th\tP^{-1}R^{\alpha}(\sigma^2+\IR) \ |\  R=r ]  r \dd r. \nonumber
\end{align*}
Using the fact that $\rh \sim \exp(1)$,   the inner probability term can be  further simplified as
\begin{align}
\PP[\rh>\Th\tP^{-1}R^{\alpha}(\sigma^2+\IR)\ |\ R= r] &= \E_{\Ir} \left[\PP[\rh>\Th\tP^{-1}R^{\alpha}(\sigma^2+\IR) \ |\  R=r, \IR] \right] \nonumber \\
    &= \E_{\Ir}\left[\exp(-\Th\tP^{-1}r^{\alpha}(\sigma^2+\Ir))  \right] \nonumber \\
    &=   e^{-\tP^{-1} \Th r^{\alpha} \sigma^2}{\cal L}_{\Ir}(\Th\tP^{-1}r^{\alpha}),
\end{align}
where ${\cal L}_{\Ir}(s)$ is the interference Laplace transform we just computed.  This gives a coverage expression
\begin{equation}
\pc(\Th,\lambda,\alpha) = 2 \pi \lambda\int_{r>0} e^{-\pi \lambda r^2}  e^{-\Th\tP^{-1}r^{\alpha} \sigma^2} {\cal L}_{\Ir}(\Th\tP^{-1}r^{\alpha})  r \dd r. \label{eq:pc11}
\end{equation}
From \eqref{eq:laplaceexp} we have
\begin{equation*}
{\cal L}_{\Ir}(\Th \tP^{-1}r^{\alpha}) = \exp\left(-2\pi \lambda \int_{r}^\infty \frac{\T}{\T+(x/r)^\alpha} x \dd x\right),
\end{equation*}
and employing a change of variables $u = \left(x/r \right)^2\Th^{-\frac{2}{\alpha}}$ results in the expression
\begin{equation}
{\cal L}_{\Ir}(\Th\tP^{-1}r^{\alpha}) = \exp\left(-\pi r^2 \lambda \rho(\Th,\alpha) \right),
\label{eq:Ir1}
\end{equation}
where
\begin{equation}
\rho(\Th,\alpha)= \Th^{2/\alpha}\int_{\Th^{-2/\alpha}}^\infty \frac{1}{1+u^{\alpha/2}} \dd u.
\label{eq:rho}
\end{equation}

The following theorem provides the final expression for the coverage probability, which is found by plugging \eqref{eq:Ir1} into \eqref{eq:pc11} and simplifying, with a final substitution of $v = r^2$.

\begin{theorem}
\label{thm:main2}
The probability of coverage of a typical randomly located mobile user is
\begin{equation}
\pc(\Th,\lambda,\alpha) = \pi \lambda \int_0^\infty  e^{-\pi \lambda v (1 + \rho(\Th,\alpha)) - \Th \snr^{-1} v^{\alpha/2}} \dd v,
\label{eq:expInt}
\end{equation}
 \end{theorem}

This fairly simple integral expression already hints at some of the key dependencies on the SINR distribution in terms of the network parameters.  However, it can be further simplified in three practical special cases that we now explore.


\subsection{Special Cases}
We now consider three special cases where the expression in Theorem 1 can be further simplified.  These correspond to exploring the high SNR regime $\snr \to 0$ -- equivalently referred to as  the ``no noise'' or ``interference-limited'' case -- and to the case where the path loss exponent is constrained to be $\alpha = 4$, which is a fairly typical value for terrestrial propagation at moderate to large distances \cite{GolBook}.  There are three such combinations of these simplifications that we consider.

\subsubsection{Noise still present, $\alpha = 4$}
 In this case, the probability of coverage can be written as
\begin{equation}
\pc(\Th,\lambda,\alpha) = \pi \lambda \int_0^\infty  e^{-\pi \lambda v (1 + \rho(\Th,4)) - \Th \tP^{-1}\sigma^2 v^{2}} \dd v,
\end{equation}
where $\rho(\Th,4)$ can be computed as
\begin{equation}
\rho(\Th,4)= \sqrt{\Th}\int_{\sqrt{\Th}}^\infty \frac{1}{1+u^{2}} \dd u=\sqrt{\Th}\arctan{\sqrt{\Th}}. 
\end{equation}
Let us define $\kappa(\Th) = 1+\rho(\Th,4)$. 

Now, note that
\begin{equation}
\int_0^\infty e^{-a x} e^{-b x^2} \dd x= \sqrt{\frac{\pi}{b}} \exp\left(\frac{a^2}{4b}\right) Q\left(\frac{a}{\sqrt{2 b}}\right),
\label{eq:exp-int}
\end{equation}
where $Q(x) = \frac{1}{\sqrt{2\pi}}\int_x^{\infty} \exp(-y^2/2)\dd y$ is the standard Gaussian tail probability. 

 Using above result, we get for $\alpha = 4$ in Theorem 1:
\begin{eqnarray}
\pc(\Th,\lambda,4) & = & \frac{\pi^{\frac{3}{2}}\lambda}{\sqrt{\Th/\snr }}  \exp\left(\frac{ (\lambda \pi \kappa(\Th))^2 }{4 \Th/\snr}\right) Q\left(\frac{\lambda \pi\kappa(\Th)}{ \sqrt{2 \Th/\snr }}\right) \label{eq:CovConstNoise},
\end{eqnarray}
where $\snr=\tP/\sigma^2$.
This expression is practically closed-form, requiring only the computation of a simple $Q(x)$ value  which is similar to the BER expression for a single link in AWGN.

\subsubsection{Interference-limited, any path loss exponent}
The coverage probability for the noiseless case can be easily obtained from Theorem \ref{thm:main2} by substituting $\snr = \infty$ and evaluating the now trivial $e^{ax}dx$ integral.  The result is given by the following simple expression
\begin{equation}
\pc(\Th,\lambda,\alpha) = \frac{1}{1 + \rho(\Th,\alpha)}.
\label{eq:SIR-1tier}
\end{equation}

\subsubsection{Interference-limited, $\alpha = 4$}
When the path loss exponent $\alpha=4$, the no noise coverage probability can be further simplified to
\begin{equation}
\pc(\Th,\lambda,4) =  \frac 1 {1+\sqrt{\Th} \arctan\sqrt{\Th}} \label{eq:CovNoNoise}.
\end{equation}
This is a remarkably simple expression for coverage probability that depends only on the SIR threshold $\Th$, and as expected it goes to 1 for $\Th \to 0$ and to 0 for $\Th \to \infty$.  For example, if $\Th = 1$ (0 dB, which would allow a maximum rate of 1 bps/Hz), the probability of coverage in this fully loaded network is $0.56$. 

Cellular engineers will notice that this value of coverage probability seems low for $\Th = \sinr = 1$.  This is for several reasons, including that we have ignored sectoring and other antenna gains and that all BSs are transmitting at all times (worst-case interference, i.e. a fully loaded frequency reuse 1 system).

\subsection{Validation}

A natural question to ask is whether these mathematical results reasonably describe real-world cellular networks, which do not typically have a Poisson BS distribution.  Real cellular networks are obviously deployed in a more strategic manner than just random independent dropping, and for this reason a regular grid -- either square or hexagonal -- has been used most frequently.  However, this is idealized in the other direction, and are too perfectly regular. Thus, for simulation-based studies the grid-based BS locations are sometimes perturbed by a random variable, for example a zero mean 2D Gaussian or a 2D uniform random variable \cite{NieLemC2003}, to account for the imperfections relative to the regular grid.

\begin{figure} [ht]
\begin{center}
\includegraphics[width=0.40\columnwidth]{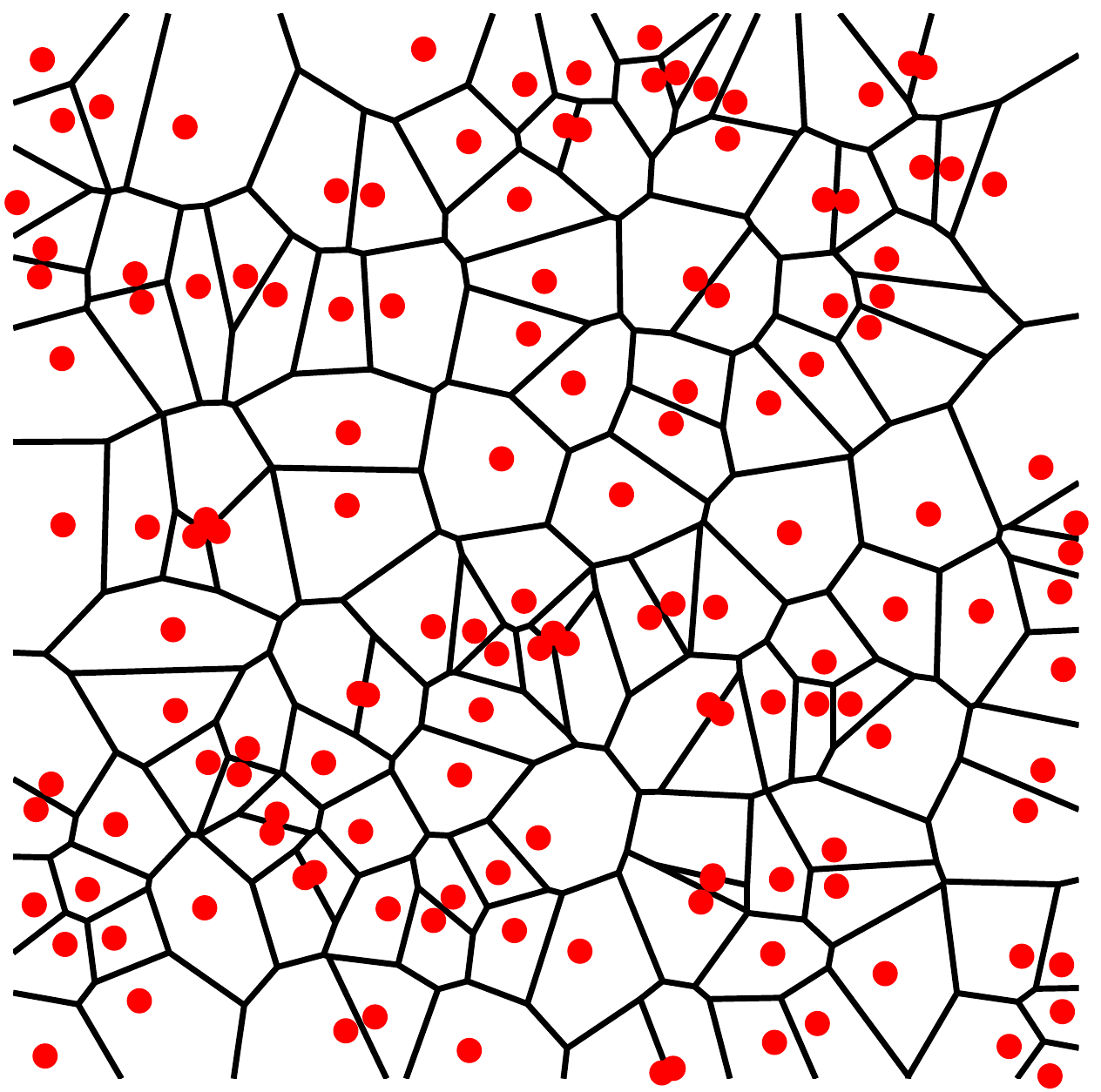}\quad \quad
\includegraphics[width=0.40\columnwidth]{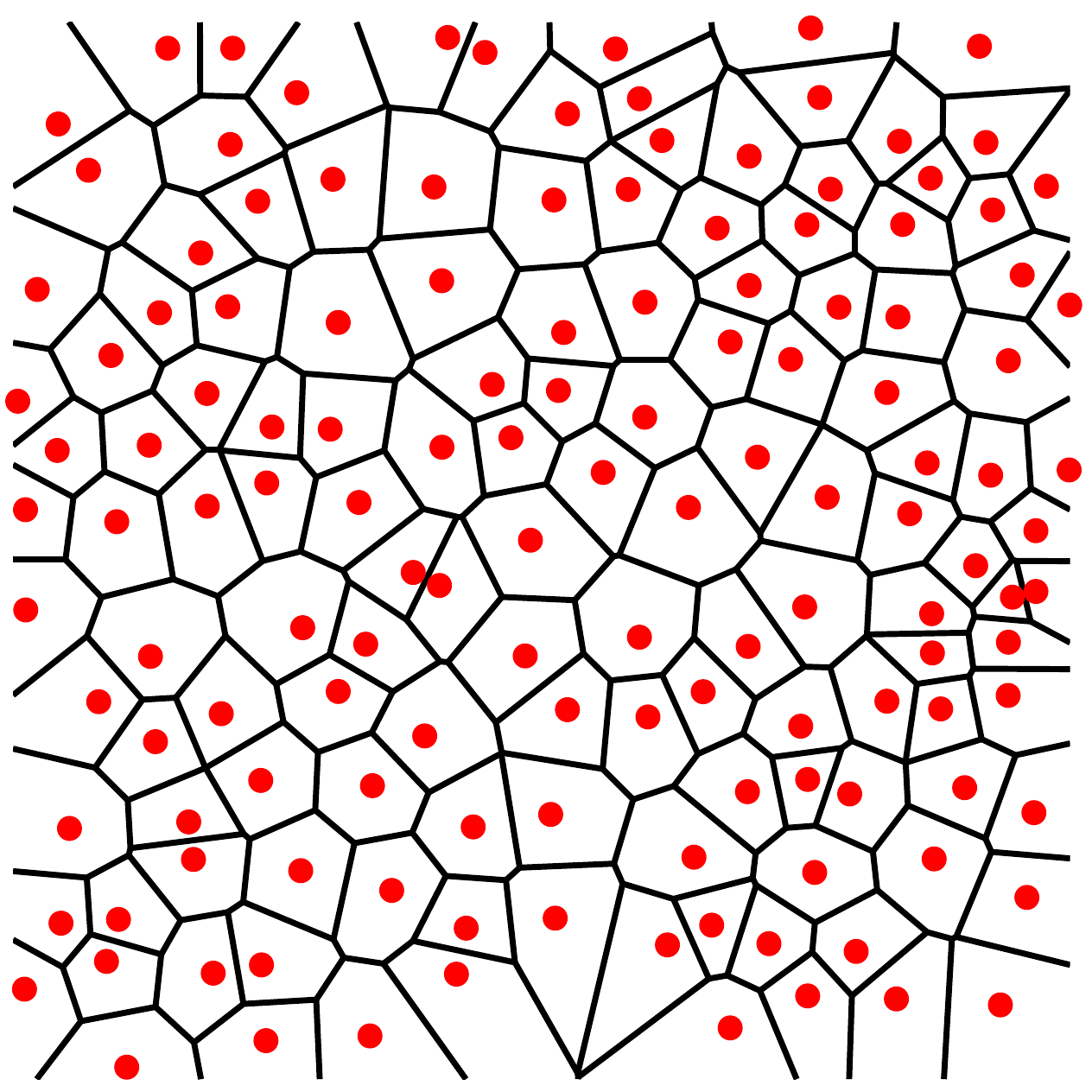}
   \caption{Poisson BSs (left) compared with Actual LTE BSs over a $40 \times 40$ km area.}
   \label{fig:layouts}
   \end{center}
\end{figure}

A representative illustration can be seen in Fig. \ref{fig:layouts} that shows a $40$ km $\times 40$ km section of a real-world LTE network in a large flat urban American city, and a sample of BSs from a Poisson point process of the same density.  One can also easily imagine a hexagonal or square grid. Subjectively, it is straightforward to observe that this real-world LTE network lies somewhere between the two extremes of perfect regularity (hexagonal or square grid) and complete randomness (PPP).  Thus, we would expect that the SINR coverage probability of a real-world LTE-like cellular network to also be bounded by these two extremes, as pointed out long ago by \cite{Bro00}. 

\begin{figure} [ht]
\begin{center}
   \includegraphics[width=0.5\columnwidth]{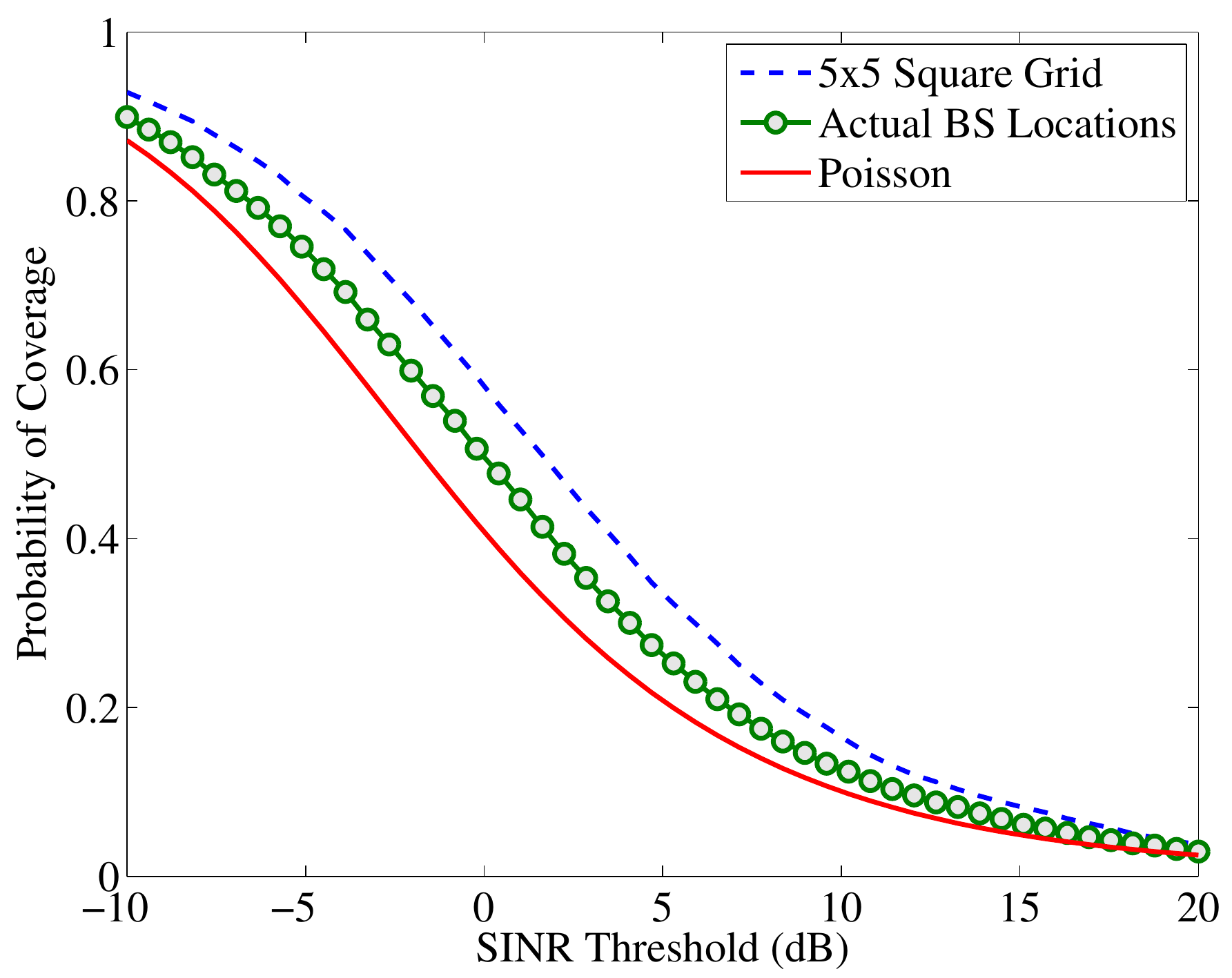}
   \caption{A comparison of the interference-limited coverage probability with $\alpha = 3$ for 3 different cellular layouts: a $5 \times 5$ square grid (simulated), an actual LTE network layout in a large urban area (simulated), and a Poisson layout, computed by \eqref{eq:SIR-1tier}}
   \label{fig:coverage-compare}
   \end{center}
\end{figure}

Indeed, the  coverage probability does quantitatively lie between these two extremes in general.  A representative plot is given in Fig. \ref{fig:coverage-compare}, where the coverage probability for the actual BS locations is seen to lie roughly between a square grid and the PPP.  The curves have the same shape and it can be observed that the Poisson curve is pessimistic by about 2dB over nearly the entire SINR range.  The gap depends on the actual BS layout used, as well as the path loss exponent.

This strong similarity between the SINR coverage probabilities for different BS layouts has been explored rigorously in subsequent work.  Haenggi and his collaborators have shown that for BSs drawn from any motion invariant point process, it is possible to rigorously characterize the SINR coverage probability using the corresponding expression for a PPP layout, and adding a horizontal SINR shift in dB \cite{Hae14,GuoHae15}.  For example, he shows that the shift from the PPP to a hexagonal lattice is 3.4 dB which is the maximum for any layout. This is similar to the shift we can empirically observe in Fig. \ref{fig:coverage-compare} for the square grid, which Haenggi showed had an average shift of exactly 3 dB \cite{Hae14}.   Therefore, one can use the PPP for analysis, and then add a calibrated horizontal SINR shift of 1-3 dB depending on the regularity of desired BS layout.  

Another interesting recent theoretical result shows that even completely regular networks like the hexagonal grid effectively appear to be Poisson when there is sufficient shadowing \cite{Blaszczyszyn2013}.  That is, instead of modeling the network as hexagonal grid BSs with lognormal shadowing, one could instead simply perturb the BS locations (and hence the path loss from each one) to account for the randomness introduced by shadowing.  With sufficiently large shadowing (about 10dB, a realistic value), the perturbations are large enough to make the resulting point process effectively Poisson.  Thus, the randomness in the PPP layout (and its resulting loss in SINR vs. a regular layout) can be interpreted as being due to lognormal shadowing, which provides further justification for using a PPP for the BS locations. The same basic idea can be used to incorporate the effect of shadowing in the downlink analysis, which we discuss next.

\subsection{Incorporating shadowing} \label{sec:shadowing}
We now discuss a simple way of incorporating the effect of shadowing in the coverage analysis. We enrich the system model described in Section \ref{sec:DLSysMod} slightly by assuming that each link from the BS $\mathbf{X}_i$ additionally suffers from shadowing with gain $\chi_i$.
We assume shadowing gains $\{\chi_i\}$ to be i.i.d. and also independent of the locations of the BSs $\{\rX_i\}$. 
Now, the instantaneous received power at the user from the $i\ths$ BS is given as
\begin{align}
p_{ri,\text{inst}}&=H_ip\|\mathbf{X}_i\|^{-\alpha}\chi_i.
\end{align}
To be consistent with the discussion so far, we consider maximum average power-based cell association rule. Since shadowing is a long-term effect, it will also appear in the association rule while deciding the serving BS for each user. Therefore, the average power received from the BS located at $\mathbf{X}_i$ is
\begin{align}
p_{ri}&=p\|\mathbf{X}_i\|^{-\alpha}\chi_i.
\end{align}
While the presence of additional random variable $\chi_i$ naturally seem to complicate the analysis, we will show that it is possible to absorb it in the location term $\mathbf{X}_i$ to define an {\em equivalent PPP} $\Phi_D$, which has the same coverage probability, but does not have any shadowing. We can then use the result derived in the previous subsections to compute the coverage probability for this equivalent PPP. More precisely, let us define a PP $\Phi_D$ as
\begin{align}
\Phi_D=\{\mathbf{Y}_i:\mathbf{Y}_i={\rX_i} \chi_i^{-1/\alpha},\rX_i\in\Phi\}.
\end{align}
It can be seen that the association, the serving power and the interference for the derived BS PP is the same as the ones for the original PPP.  Therefore, the coverage probability of the derived  PP is the same as that of the original PPP. From the displacement theorem \cite{BacNOW}, we know that the derived PP is a PPP with expectation measure given as
\begin{align}
\mu_D(\setnot{A})&=\int_{\mathbb{R}^2}\lambda \P\left[\x\chi^{-1/\alpha}\in \setnot{A}\right]\dd\x \nonumber\\
&=\int_{\mathbb{R}^2}\lambda \E\left[\ind{\x\chi^{-1/\alpha}\in \setnot{A}}\right]\dd\x \nonumber\\
&=\lambda \E\left[\int_{\mathbb{R}^2}\ind{\x\chi^{-1/\alpha}\in \setnot{A}}\dd\x\right] .
\end{align}

Now, let $\setnot{A}=\{\mathbf{z}:\|\mathbf{z}\|\le r\}$, then
\begin{align}
\mu_D(\setnot{A})
&=\lambda \E\left[\int_{\mathbb{R}^2}\ind{\|\x\|\le r\chi^{1/\alpha} }\dd\x\right] \\
&=\lambda \E\left[\pi r^2\chi^{2/\alpha} \right]=\lambda \pi r^2 \E\left[\chi^{2/\alpha} \right].
\end{align}
The necessary and sufficient condition for the last term in the above equation to exist is that $\E\left[\chi^{2/\alpha} \right]<\infty$ \cite{DhiAndJ2014}. Now the density of the derived PPP can be obtained from the expectation measure as
\begin{align}
\lambda_D(r)
&=\frac{1}{2\pi r}\frac{\dd}{\dd r }\mu(\setnot{A})= \lambda\E\left[\chi^{2/\alpha} \right].
\end{align}
Therefore, the coverage probability of the derived PPP can by obtained by \eqref{eq:expInt} with $\lambda$ replaced by $\lambda_D$ which is also the coverage probability of the original PPP. Interested reader is advised to refer to the following representative set of works that use this general idea:~\cite{Blaszczyszyn2013,DhiAndJ2014,Brown2014,MadResJ2016,BlaKeeC2013,KeeBlaC2013,ZhaHaeJ2014}. 

\newcommand{\UE}{\mathrm{U}_0}
\newcommand{\PhiU}{\Phi_\mathrm{a}}

\section{Uplink Analysis}
\label{sec:UL}

We now consider the uplink analysis of a cellular network as first explored in \cite{NovDhi13} and later extended in several followup publications such as~\cite{SinZhaJ2015,YuMukC2012,ElSHosJ2014,LeeSanJ2014,DiGuaJ2016,MarGomJ2016}. We will provide more details on these works at the end of this section. Similar to the downlink case, the uplink analysis focuses on the distribution of the received $\sinr$ from the typical mobile user at its {\em serving} BS (termed {\em tagged} BS), with the interference now coming from all the mobile users in the rest of the network. Fig. \ref{fig:sysMod} gives a visual representation of the uplink system model. Unlike the downlink case, the exact characterization of the uplink $\sinr$ is not available. The main goal of this section is to highlight key challenges that appear in the analysis and provide one representative analytical approach that gives a close approximation for the $\sinr$ distribution. Pointers to other approaches (usually variants of the presented approach) will also be provided at the end of the section.

\subsection{Model and Preliminaries}

Many of the downlink modeling assumptions from Sect. \ref{sec:DL} are adopted here, including the BS locations following a homogeneous PPP with density $\lambda$; the use of power law path loss, Rayleigh fading, and Gaussian noise $\sigma^2$; and the maximum average received power based association rule (a mobile user would attach to the BS to which it has the smallest average path loss).

\begin{figure} [t]
\begin{center}
   \includegraphics[trim=0 210 0 0,clip,width=0.5\columnwidth]{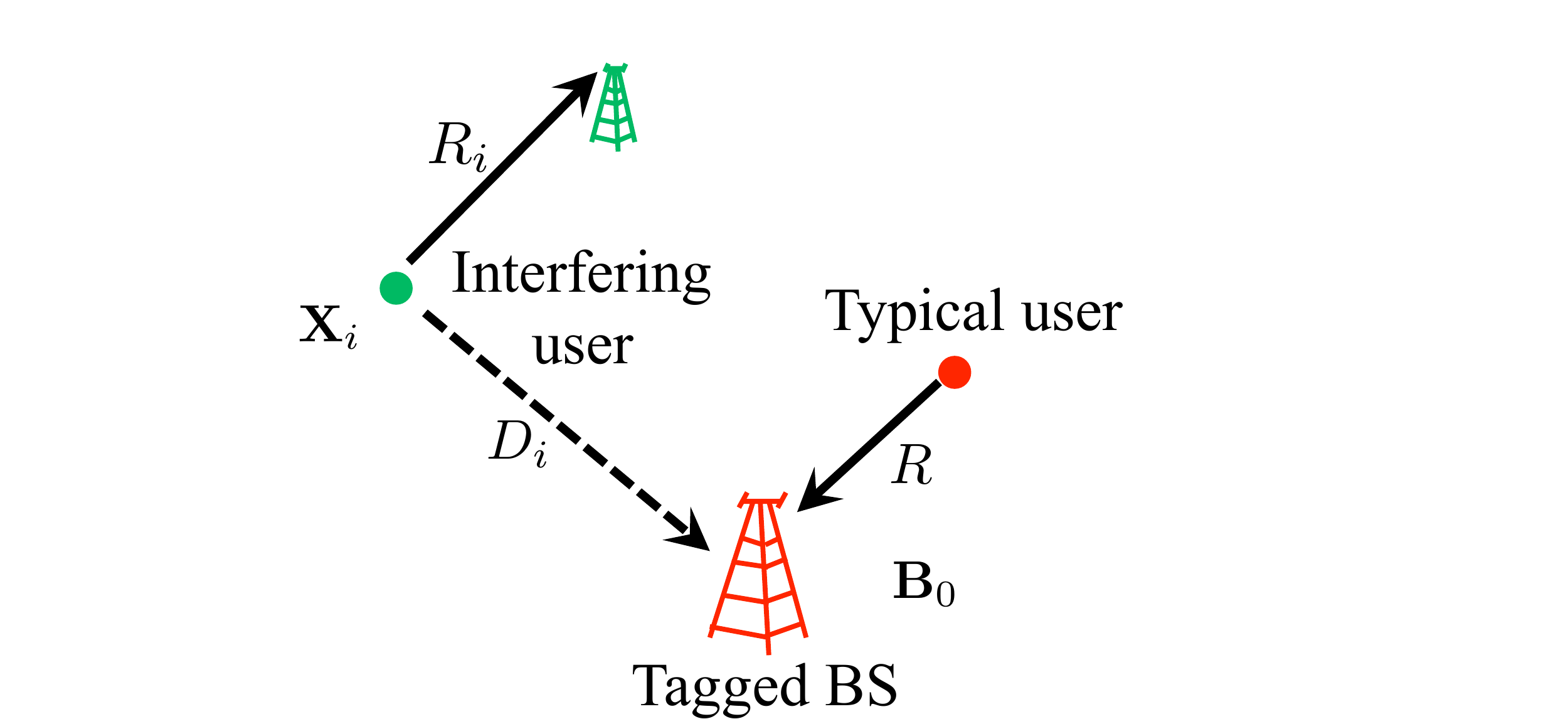}
   \caption{Uplink system model focusing on the uplink performance of the typical user served by its {\em tagged} BS located at $\rUz$.
}
   \label{fig:sysMod}
   \end{center}
\end{figure}

Additional modeling assumptions for the uplink analysis are:

\begin{itemize}
\item The mobile user locations (before associating with a particular BS) are assumed to form a realization of a homogeneous PPP with density $\lambda_\mathrm{u}$.  This is well motivated, since mobile users really do take up nearly independent locations in many circumstances.  This is how cellular networks are generally simulated in practice, by using a uniform distribution of users (which is equivalent to a PPP in a given area conditioned on the number of users).   

\item Each BS has a single active uplink user scheduled on a given time-frequency resource which is randomly chosen from all the users located in its Voronoi cell,  as in the downlink.  Therefore, the user PPP $\lambda_\mathrm{u}$ can be thinned to get a point process $\PhiU$ which denotes the locations of the active users. Crucially, this thinning is \emph{not independent} as only one user per BS can be selected among all users located in that BS's Voronoi cell. This fact causes major complications for the uplink analysis.  

\item In light of the previous point, for tractability, we assume that the active users also form a PPP even after associating just one per BS. It has been shown through simulations that this approximation, although conceptually dubious, does not effect the coverage probability result too much~\cite{NovDhi13,SinZhaJ2015}. This is also demonstrated in Fig.~\ref{fig:UL-Coverage}, where the analytical and simulation results are shown to match closely.
Since there is one active user per cell, the density $\lambda_a$ of this thinned PPP of active users should be set equal to the density of BSs, i.e. $  \lambda_a$ should be set equal to $\lambda$.

\item The typical mobile user is located at the origin and connected to the nearest BS located at $\rUz$ that we term as the {\it tagged} BS.

\item As shown in Fig. \ref{fig:sysMod}, we denote the distance between an interfering user  (located at $\rX_i$) to the tagged BS by $\|\rX_i - \rUz\|=D_i$, and the distance of the interfering mobile to its serving BS by $R_i$. The distance between the tagged BS and the typical user (for which the uplink $\sinr$ distribution is computed) is denoted by a random variable $R$.

\item The mobiles utilize distance-proportional fractional power control of the form $R_i^{\alpha \epsilon}$, where $\epsilon \in [0, 1]$ is the power control factor. Thus, as a user moves closer to the desired BS, the transmit power required to maintain the same received signal power decreases.  This is very similar to the uplink operation in LTE and includes fixed transmit power ($\epsilon = 0$) and perfect channel inversion ($\epsilon = 1$) as special cases. 

\end{itemize}

Unlike the downlink case, where the interference field could be readily modeled as a homogeneous PPP outside an exclusion zone defined by the serving BS, the characterization of the interference field in the uplink case is far more challenging. For instance, note that an interferer (mobile user) in the uplink case can actually lie closer to the tagged BS than the transmitter of interest (typical user). One way of approximately characterizing the interference field is by accounting for the simple fact that a user is deemed to be interfering if it is not served by the tagged BS (our BS of interest). If we denote the distance of an interferer to the tagged BS by $d$, this event is equivalent to finding at least one BS in a ball of radius $d$ centered at the interfering user. The probability of this event is therefore given by $(1-\exp(-\pi \lambda_0 d^2))$. In other words, a user located at the distance $d$ from the tagged BS will be deemed interfering with probability $(1-\exp(-\pi \lambda_0 d^2))$. Therefore, the effective interference field as observed from the tagged BS can be modeled as a non-homogeneous PPP $\Phi_{Ia}$ with a radially symmetric distance dependent intensity function $\lambda_{Ia}(d) = \lambda(1 - \exp(-\pi \lambda d^2))$ (relative to the tagged BS) \cite{SinZhaJ2015}.


Under this system model, the uplink $\sinr$ of the typical user at the tagged BS
is
\begin{equation}
\label{eq:SINR}
\sinr = \frac{\rh \tP R^{\alpha\left(\epsilon - 1\right)}}{{\sigma}^2 + I},
\end{equation}
where interference $I$ is given by
\begin{equation}
\label{eq:IRUP}
I = \displaystyle\sum_{\rX_i\in \Phi_{Ia}}{ \tP\left(R_i^{\alpha}\right)^{\epsilon}\rg_i{\|\rX_i - \rUz\|}^{-\alpha}}.
\end{equation}
If $\epsilon = 1$, the numerator of \eqref{eq:SINR} becomes $\rh\tP$, with the path loss completely inverted by power control, and if $\epsilon = 0$ no channel inversion is performed and all the mobiles transmit with the same power $\tP$. Recall also from the above discussion that the interference field $\Phi_{Ia}$ and hence the interference power distribution turn out to be  independent of the desired signal power from the typical user (given by the numerator term $\rh \tP R^{\alpha\left(\epsilon - 1\right)}$ of \eqref{eq:SINR}). This is an outcome of the way interference field is modeled and will be useful in the coverage probability analysis presented later in this section.

\subsection{Distribution of Serving Link Distances ($R$ and $R_i$)}
For the above setup, the random variable $R$ can be shown to be Rayleigh distributed by differentiating the null probability of PPP  similar to the previous section. Its PDF is given by
\begin{equation}
f_{R}(r) = 2\pi\lambda r e^{-\lambda \pi r^2},\ r \geq 0,
\label{eq:Rpdf}
\end{equation}
which for this setup is {\em exact}.


Recall that $R_i$ is the link distance between the interfering user and its serving BS, which is determined by the {\em smallest average path loss} rule. Now, the random variables $\{R_i\}$ are identically distributed but not independent in general. The dependence is induced by the structure of the Poisson-Voronoi tessellation and the restriction that only one user can be active in each Voronoi cell in the time-frequency resource block of our interest. A reasonable approximation for the marginal distribution of $R_i$ is the Rayleigh distribution (given by \eqref{eq:Rpdf}).\footnote{As an aside, it should be noted that \eqref{eq:Rpdf} slightly overestimates $R_i$ because of the fact that the tagged cell is on an average larger than a randomly chosen cell in the network, which means $R_i$ should in general be smaller than $R$. This disparity is because of the {\em waiting bus paradox} (also called Feller's paradox). Interested readers should refer to~\cite{YuMukC2012} for more details.}
As discussed earlier, modeling dependencies between these random variables is however very challenging. That being said, the dependence between $R_i$ and the distance $D_i$ of the interfering user from the tagged BS can be approximately modeled by noticing that $R_i$ is always upper bounded by $D_i$. Note that $R_i$ cannot be bigger than $D_i$ because the distance of an interfering user to its serving BS has to be smaller than its distance to the tagged BS (otherwise tagged BS will become its serving BS). Accounting for this fact, the 
distribution of $R_i$ conditioned on $D_i$ is given by 
\begin{equation}
f_{R_i}(r|D_i) =  \frac{2\pi \lambda r \exp(-\lambda \pi r^2)}{1-\exp(-\pi \lambda D_i^2)}\ , 0 \leq r \leq D_i,
\label{eq:pdfRz}
\end{equation}
which is a truncated version of the Rayleigh distribution. As noted earlier, the above distribution is an approximation. Using these distance distributions, we characterize interference next.

\subsection{Interference Characterization}
The net interference at the tagged BS is the sum of powers from all the transmitting mobiles (modeled by the PPP $\Phi_{Ia}$). Under the power control model described in the previous subsection, this power depends upon the distance of a mobile to its serving BS and the power control factor $\epsilon \in [0,1]$. For this setup, we now compute the Laplace transform of interference distribution observed at the tagged BS. Recalling that the interference is denoted by $I$, we get:
\begin{align}
\mathcal{L}_{I}(s) &= \E_I\left[e^{-sI} \right]
= \E_{I}\left[\exp\left(-\sum_{\rX_i \in \Phi_{Ia}} s\tP R_i^{\alpha \epsilon}\rg_i\|\rX_i - \rUz\|^{-\alpha}\right)\right]\\
&= \E_{R_i,\rg_i,\rX_i}\left[\prod_{\rX_i \in \Phi_{Ia}}\exp\left(-s \tP R_i^{\alpha \epsilon} \rg_i\|\rX_i - \rUz\|^{-\alpha}\right)\right].
\end{align}
Using the independence of $\rg_i$'s and $R_i$'s across $\Phi_{Ia}$, we get
\begin{align}
\mathcal{L}_{I}(s)&=\E_{\Phi_{Ia}}\left[\prod_{\rX_i \in \Phi_{Ia}}\E_{R_i,\rg_i}\left[\exp\left(-s\tP R_i^{\alpha \epsilon} \rg_i\|\rX_i - \rUz\|^{-\alpha}\right)\right]\right].
\end{align}
Since $\rg_i\sim \exp(1)$, 
\begin{align}
\mathcal{L}_{I}(s)= \E_{\Phi_{Ia}}\left[\prod_{\rX_i \in \Phi_{Ia}}\E_{R_i}\left[\frac{1}{1+ s\tP R_i^{\alpha \epsilon} \|\rX_i - \rUz\|^{-\alpha}}\right]\right]. \end{align}
Now using the PGFL of a PPP and the fact that the interference field $\Phi_{Ia}$ is a non-homogeneous PPP with distance-dependent density function $\lambda_{Ia}(d) = \lambda (1-\exp(-\pi \lambda d^2))$ relative to the tagged BS, we get 
\begin{align}
\mathcal{L}_{I}(s)&= \exp\left(-2\pi\lambda\int_{0}^{\infty} (1-\exp(-\pi \lambda x^2))\left(1 - \E_{R_i}\left[\frac{1}{1+ s\tP R_i^{\alpha \epsilon} x^{-\alpha}}\right] \right)x \dd x\right)\nonumber\\
&= \exp\left(-2\pi\lambda\int_{0}^{\infty}(1-\exp(-\pi \lambda x^2))
\E_{R_i}\left[\frac{1}{1+ (s\tP)^{-1} R_i^{-\alpha \epsilon} x^{\alpha}}\right] x \dd x\right)
\label{eq:laplace2}.
\end{align}
Using the PDF  of $R_i$, given by \eqref{eq:pdfRz}, the Laplace transform of interference can be further simplified to
\begin{align}
\mathcal{L}_{I}(s) =\exp\left(-2\pi\lambda\int\limits_{0}^{\infty}
\int\limits_{0}^{x^2} \frac{1}{1 + (s \tP)^{-1}u^{-\alpha \epsilon/2} x^{\alpha}} \pi \lambda e^{-\lambda \pi u} \dd u 
x \dd x\right).
\label{eq:laplacePPP}
\end{align}
Recall that in this approach the interference distribution $I$ is not a function of the distance $R$ from the typical user to its tagged BS. This is the key difference between the downlink analysis presented in the previous section and the approximate uplink analysis presented in this section.

\subsection{Coverage Probability \label{sec:cp}}
Similar to previous section, we will compute the uplink coverage probability which is defined as the CCDF of uplink $\sinr$, i.e., the probability that the uplink $\sinr$ at the tagged BS is greater than the target $\sinr$ $\Th$. 
Starting from the definition of uplink coverage probability and $\sinr$, we get
\begin{align}
\label{eq:maineq}
\pc(\T,\lambda,\alpha,\epsilon)&= \int_0^\infty \PP\left(\sinr > \T\right)f_R(r) \dd r\\
&= \int_0^\infty \PP\left(\frac{\rh \tP\left(r^{\alpha(\epsilon-1)}\right)}{{\sigma}^2 + I} > \T\right)2\pi\lambda r e^{-\pi\lambda r^2} \dd r\\
&= \int_0^\infty \PP\left( \rh> \frac{\T({\sigma}^2 + I)}{\tP r^{\alpha(\epsilon-1)} }\right)2\pi\lambda r e^{-\pi\lambda r^2} \dd r\end{align}
Since $\rh\sim \exp(1)$, we get
\begin{align}
\pc(\T,\lambda,\alpha,\epsilon)& = \int_0^\infty 2\pi\lambda r e^{-\pi\lambda r^2}e^{- \T\tP^{-1}r^{\alpha(1-\epsilon)}\sigma^2}\E_{I}\left[e^{-\T\tP^{-1}r^{\alpha(1-\epsilon)}I}\right] \dd r\\
&= \int_0^\infty 2\pi\lambda r e^{-\pi\lambda r^2}e^{-\T\tP^{-1}r^{\alpha(1-\epsilon)}\sigma^2}\mathcal{L}_{I}\left(\T\tP^{-1}r^{\alpha(1-\epsilon)}\right) \dd r.
\end{align}
Now the Laplace transform of interference at $s=\T\tP^{-1}r^{\alpha(1-\epsilon)}$ is given as
\begin{align}
&\mathcal{L}_{I}(\T\tP^{-1}r^{\alpha(1-\epsilon)}) =\exp\left(-2\pi\lambda\int\limits_{0}^{\infty}
\int\limits_{0}^{x^2} \frac{1}{1 + {\T}^{-1}r^{-\alpha(1-\epsilon)} u^{-\alpha \epsilon/2} x^{\alpha}} \pi \lambda e^{-\lambda \pi u} \dd u 
x \dd x\right).
\label{eq:laplacePPPwithvalue}
\end{align}

 Using the expression of Laplace transform of interference, we now provide the final coverage expression in  Theorem~\ref{thm:uplink1tier}. 


\begin{theorem}
\label{thm:uplink1tier}
The uplink coverage probability is given by:
\begin{align}
 \pc(\T,\lambda,\alpha,\epsilon)  = 2\pi\lambda \int_0^\infty r e^{-\pi\lambda r^2-
  \T\tP^{-1} r^{\alpha(1-\epsilon)}\sigma^2}\nu(r,\lambda,\alpha,\epsilon) \dd r,
\label{eq:ulMain}
\end{align}
where $\nu(r,\lambda,\alpha,\epsilon)$ is given by 
\begin{align}
&\nu(r,\lambda,\alpha,\epsilon)=\exp\left(-2\pi\lambda\int\limits_{0}^{\infty}
\int\limits_{0}^{x^2} \frac{1}{1 + \T^{-1}r^{-\alpha(1-\epsilon)} u^{-\alpha \epsilon/2} x^{\alpha}} \pi \lambda e^{-\lambda \pi u} \dd u 
x \dd x\right).
\label{eq:uplink}
\end{align}
\end{theorem}

\begin{figure} [t]
\begin{center}
   \includegraphics[width=0.5\columnwidth]{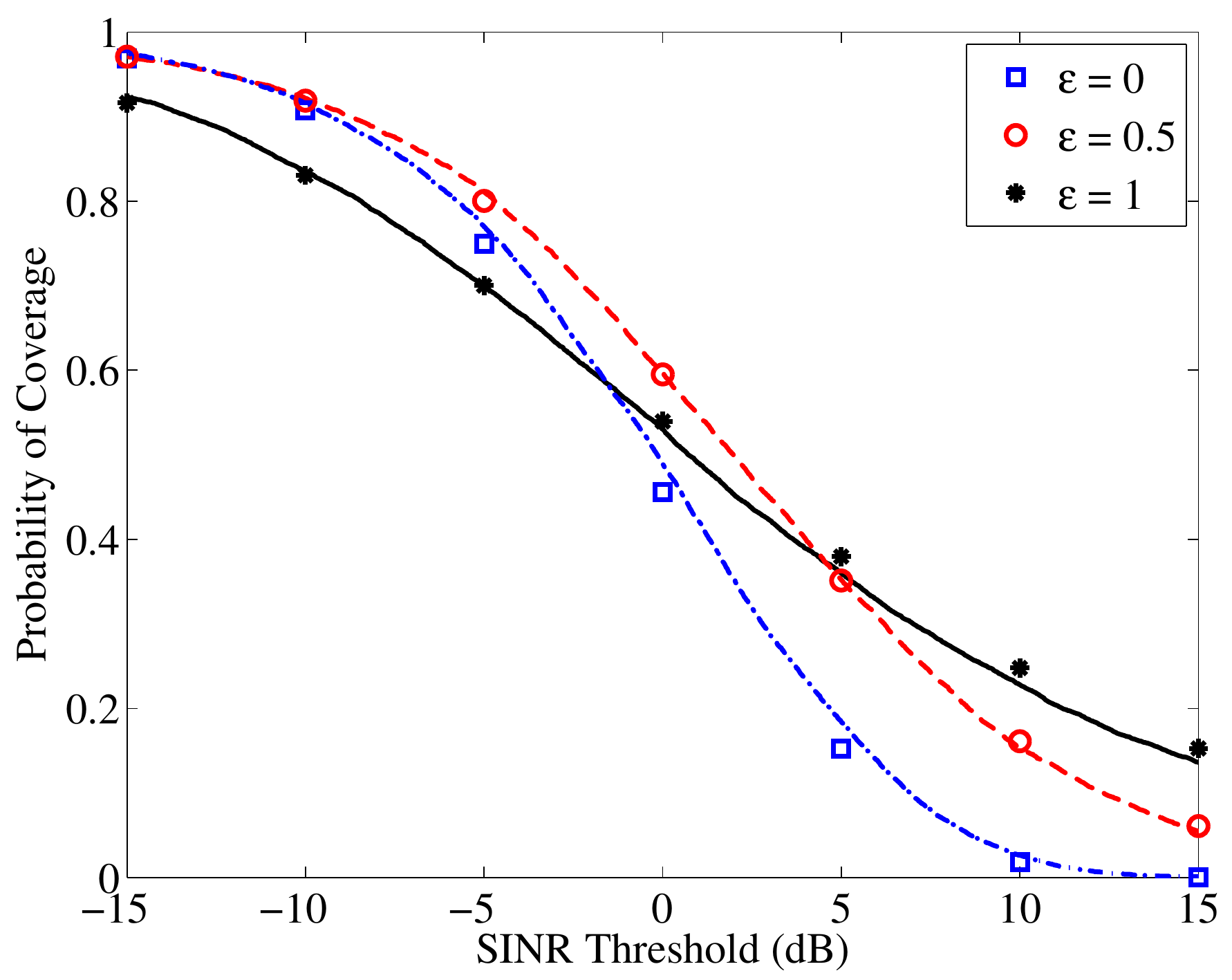}
   \caption{Uplink coverage probability of a typical user for several values of the power control fraction $\epsilon$. Lines and markers correspond to the simulation and theoretical results, respectively.}
   \label{fig:UL-Coverage}
   \end{center}
\end{figure}

Since this result is derived under various simplifying assumptions, it is important to validate it by comparing it with the results obtained from Monte Carlo simulations. We perform this comparison in Fig.~\ref{fig:UL-Coverage} for the setup with $\lambda = 4 \times 10^{-6}$ BS/m$^2$ (4 BSs per square km), $\alpha = 4$, and $\tP = 1$. We assume interference-limited scenario, where the interference power dominates the thermal noise (i.e., $\sigma$ can be assumed to be $0$). We get an almost perfect match for all values of the power control parameter $\epsilon$. Through extensive simulations, we have noticed that this expression works quite well for all the values of simulation parameters that we considered. This is attributed to the careful handling of dependencies between various random variables involved in the derivation. 


\subsection{Special Case}
The coverage probability expression can be simplified for the full channel inversion power control case ($\epsilon=1$) in the interference-limited scenario. The uplink coverage probability for the full power control case ($\epsilon = 1$) assuming no noise ($\sigma^2=0$) is given by
\begin{equation}
\pc(\T,\lambda,\alpha,\epsilon=1) = \int_0^\infty 2\pi\lambda r e^{-\pi\lambda r^2}\mathcal{L}_{I}\left(\T\tP^{-1}\right) \dd r
\end{equation}
where $\mathcal{L}_{I}(s)$  is given by \eqref{eq:laplacePPP} with $\epsilon=1$. Note that argument of $\mathcal{L}_{I}(s)$ is independent of $r$ and can hence be moved out of the integral to get
\begin{equation}
\pc(\T,\lambda,\alpha,\epsilon=1) = \mathcal{L}_{I}\left(\T\tP^{-1}\right) \int_0^\infty 2\pi\lambda r e^{-\pi\lambda r^2}\dd r=\mathcal{L}_{I}\left(\T\tP^{-1}\right) .\label{eq:fullpowercontrolpc}
\end{equation}

 Now,  \eqref{eq:laplacePPP} can be rewritten using the indicator function as follows 
\begin{align*}
\mathcal{L}_{I}(s)& =\exp\left(-2\pi\lambda\int\limits_{0}^{\infty}
\int\limits_{0}^{\infty} 1(u<x^2)\frac{1}{1 + (s \tP)^{-1}u^{-\alpha/2} x^{\alpha}} \pi \lambda e^{-\lambda \pi u} \dd u 
x \dd x\right).
\end{align*}
Interchanging the order of integration, we get
\begin{align*}
\mathcal{L}_{I}(s)&=\exp\left(-2\pi\lambda\int\limits_{0}^{\infty}
\int\limits_{0}^{\infty}\frac{ 1(u<x^2)}{1 + (s \tP)^{-1}u^{-\alpha/2} x^{\alpha}}  x \dd x  \lambda \pi e^{-\lambda \pi u}
 \dd u\right).
 \end{align*}
 Now using the variable substitution $x=u^{1/2}v$ and solving the inner integral we get
 \begin{align*}
\mathcal{L}_{I}(s)&=\exp\left(-2\pi\lambda\int\limits_{0}^{\infty}
\int\limits_{0}^{\infty}\frac{ 1(1<v)}{1 + (s \tP)^{-1} v^{\alpha}}  v \dd v  \lambda \pi u  e^{-\lambda \pi u}
 \dd u\right)\\
 &=\exp\left(-2\pi\lambda
\int\limits_{1}^{\infty}\frac{ 1}{1 + (s \tP)^{-1} v^{\alpha}}  v \dd v  \int\limits_{0}^{\infty}\lambda \pi u  e^{-\lambda \pi u}
 \dd u\right)\\
  &=\exp\left(-2\pi\lambda
\int\limits_{1}^{\infty}\frac{ 1}{1 + (s \tP)^{-1} v^{\alpha}}  v \dd v  \frac1{\lambda\pi}
\right).
\end{align*}
Using the value of $\mathcal{L}_{I}(s)$  in \eqref{eq:fullpowercontrolpc}, we get the probability coverage for the full power control as follows
\begin{align*}
\pc(\T,\lambda,\alpha,\epsilon=1)&=\mathcal{L}_{I}(\tau \tP^{-1})=\exp\left(-2
\int\limits_{1}^{\infty}\frac{ 1}{1 + (\tau)^{-1} v^{\alpha}}  v \dd v  \right)\\
&=
\exp\left(-\tau^{2/\alpha}\int_{\tau^{-1/\alpha}}^{\infty}\frac{ 1}{1 +  v^{\alpha/2}}   \dd v  \right)\\
&=\exp\left(-\rho(\tau,\alpha)\right)
\end{align*}
where last step is due to the definition of $\rho(\tau,\alpha)$ in \eqref{eq:rho}.
 Now we can compare probability of coverage for uplink full power control and downlink  for interference limited case. The downlink probability of coverage which is given by \eqref{eq:SIR-1tier}, decreases as $1/(1+\rho(\tau,\alpha))$ with $\tau$ while uplink one decreases as $\exp\left(-\rho(\tau,\alpha)\right)$ with $\tau$.
The fall off is  faster for the uplink  case due to the fact that  it is basically impossible to get a very high uplink SIR with full uplink power control.


This concludes our discussion of a representative analytical approach for uplink coverage analysis that maintains a fine balance between analytical tractability and the accurate modeling of various dependencies that appear in the uplink model. Our exposition closely followed that of \cite{SinZhaJ2015}, which provided one of the most complete analyses in this line of work. Due to the complexity involved in the uplink analysis, there are several other approaches and {\em generative models} proposed in the literature. Some of these approaches are variants of the approach discussed above. A representative set of such works is~\cite{ElSHosJ2014,LeeSanJ2014,YuMukC2012,DiGuaJ2016,MarGomJ2016}. While all these works include distance-dependent thinning discussed above, the mathematical formulation and assumptions are slightly different in each work: \cite{ElSHosJ2014} is focused on the special case of channel inversion-based power control, \cite{LeeSanJ2014} incorporates distance-dependent thinning by curve fitting,  \cite{YuMukC2012} evaluates coverage probability of a time-division duplex (TDD) based two-tier network, \cite{DiGuaJ2016} studies performance of a multiple-input multiple-output (MIMO) uplink under maximal ratio combining (MRC) and optimum combining (OC) in an HCN, and  \cite{MarGomJ2016} studies the performance of an interference aware power control mechanism with the objective of minimizing the interference to the second nearest BS for an HCN uplink scenario. The HCNs are discussed in detail next.

\section{Heterogeneous Cellular Network Analysis} \label{sec:HCN}

We now focus on the modeling and analysis of downlink heterogeneous cellular network (HCN), commonly referred to as a ``HetNet'' \cite{And13}. We consider $\K$ overlaid tiers of BSs that differ in terms of the transmit power, deployment density, and target $\sinr$.  These tiers model overlaid macro, micro, pico, and femtocells, as well as distributed antenna systems. As was the case in the previous two sections, our emphasis will be on exposing the analytical tools without worrying about covering all possible generalizations. 

\subsection{HCN Model}

The key aspects of the downlink HCN model follow \cite{DhiGan12,JoSan12} and are:

\begin{itemize}

\item The BS locations of the $i^{th}$ tier are modeled by an independent homogeneous PPP $\Phi_i$ with density $\lambda_i$. This extends the cellular model introduced in Section \ref{sec:DL} to $\K$ overlaid tiers of BSs.  

\item All the BSs of the $i^{th}$ tier are assumed to transmit at the same power $\tP_i$. As an example, we could have $\K=3$ with $\lambda_3 = 10 \lambda_2 = 100 \lambda_1$ and $\tP_1 = 10 \tP_2 = 100 \tP_3$ as an approximation for a conventional macrocell network (tier 1) overlaid with picocells (tier 2) and numerous end-user deployed low-power femtocells (tier 3).

\item To maintain generality, the target $\sinr$ for the $i^{th}$ tier is modeled as $\T_i$. Combining this with the previous two bullets, $i^{th}$ tier can be completely characterized by the tuple $\{\lambda_i, \tP_i, \T_i\}$.

\item We consider open access network where a given mobile user is allowed to connect to any BS in the network without any restriction.  This is best-case from an $\sinr$ coverage probability point of view.

\item As in Section \ref{sec:DL}, the users are modeled according to some stationary point process $\Phi_\user$, which is assumed to be independent of the BS locations. 

\item Power law path loss and i.i.d. Rayleigh fading are assumed as before.  Therefore, the received power from an $i^{th}$ tier BS distance $r$ away is $\tP_i \rh r^{-\alpha}$, with $\rh \sim \exp(1)$. Shadowing can be incorporated in the same way as discussed in Section~\ref{sec:shadowing}.

\item The analysis will again be performed on a typical mobile user assumed to be located at the origin. As in Section \ref{sec:DL}, all the BSs in the network are assumed to be active. 
\end{itemize}

In view of the above model, the $\sinr$ assuming the typical user connects to an $i^{th}$ tier BS located at $\rX_i \in \Phi_i$ is
\begin{equation}
\sinr(\rX_i) = \frac{\tP_i \rh_{\rX_i} \|\rX_i\|^{-\alpha}}{\sum_{j=1}^{\K} \sum_{\rX \in \Phi_j \setminus \rX_i} \tP_j \rh_{\rX} \|\rX\|^{-\alpha}+\sigma^2} = \frac{\tP_i \rh_{\rX_i} \|\rX_i\|^{-\alpha}}{I(\Phi \setminus \{\rX_i\})+ \sigma^2},
\label{eq:SINR_HCN}
\end{equation}
where $I(\Phi \setminus \{\rX_i\})$ denotes interference and $\sigma^2$ is the constant additive noise power.

\begin{figure}[t]
\centering
\includegraphics[width=0.45\columnwidth]{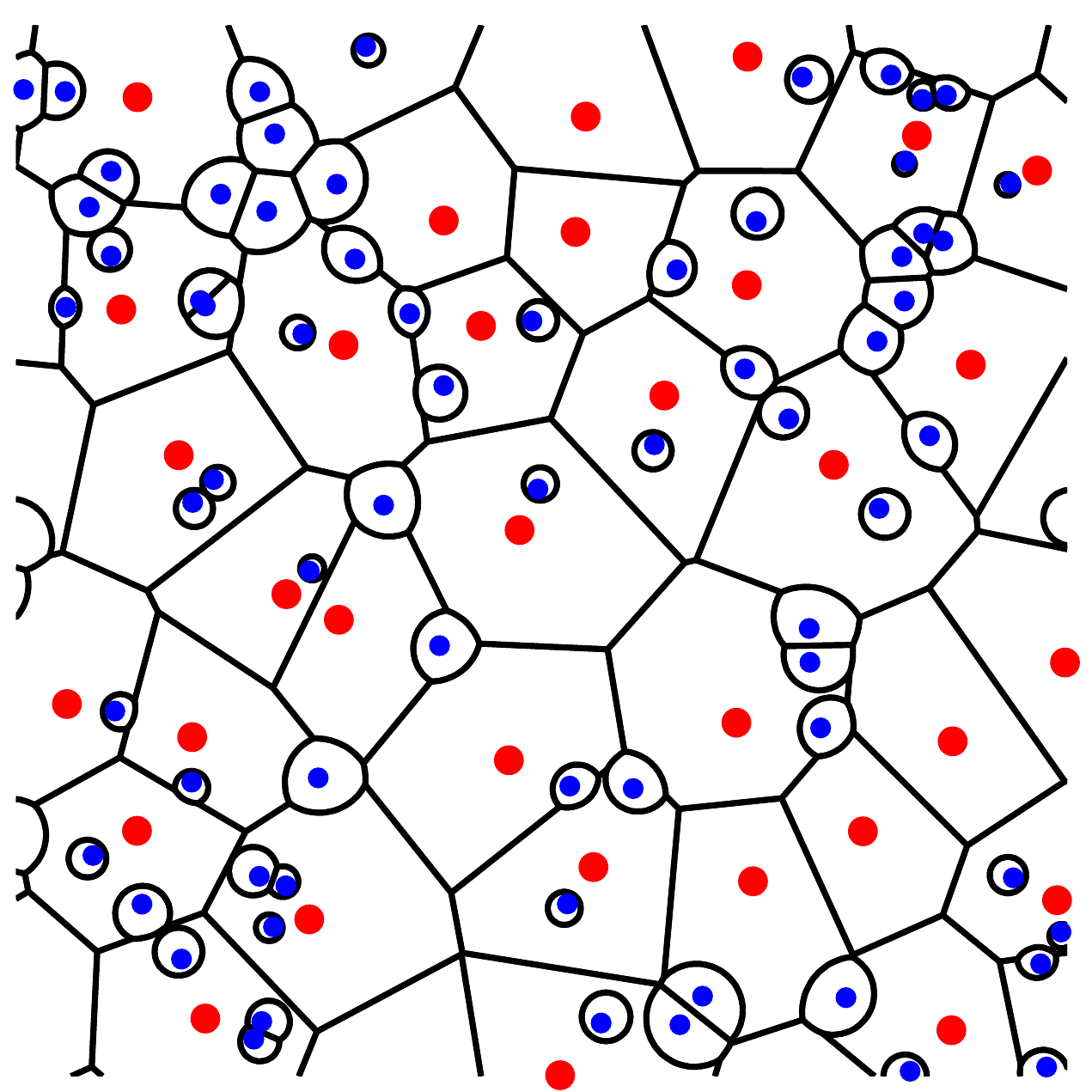}
\caption{Coverage regions in a two-tier network where macro BS locations (large circles) correspond to an actual $4$G deployment, whereas tier-2 femto (small blue sircles) BSs are modeled as independent PPPs. }
\label{fig:2HCN_RealData}
\end{figure}

\subsection{Cell Association}

In the previous two Sections, we used an the average power-based cell association, which was equivalent to associating with the closest BS in terms of the Euclidean distance. This meant that the coverage footprint of each BS corresponded to a ``cell'' of the standard Poisson-Voronoi tessellation as demonstrated in Fig.~\ref{fig:layouts}. This however is not true for HCNs due to the differences in the transmit powers of the BSs across tiers. For instance, the downlink signal from a femtocell located $100$m from a mobile user will usually be much weaker compared to the downlink signal from a macrocell located at $500$m, due to the much lower transmit power of a femtocell. This is illustrated in Fig.~\ref{fig:2HCN_RealData}, where the space is tessellated based on the maximum average received power, i.e., all the points lying in a ``cell'' receive maximum average power from the BS located at the nucleus if that cell. Due to the differences in the transmit power, this tessellation corresponds to a multiplicatively weighted Voronoi diagram, where {\em smaller cells} represent coverage footprints of low power BSs, such as femtocells. In particular, Fig.~\ref{fig:2HCN_RealData} illustrates two tier scenario where large circles represent macrocells and small blue circles represent femtocells. Clearly, the footprints of the femtocells are much smaller than those of the macrocells.


For cell selection in HCNs, we will consider two criteria, which are both quite popular in the literature. We will first consider the average power-based cell association (as in Section~\ref{sec:DL}) which was discussed in \cite{JoSan12}.
We then consider a different cell section criteria, termed as instantaneous power-based cell selection, where each user connects to its strongest BS {\em instantaneously}, i.e., the BS that offers the highest received $\sinr$ \cite{DhiGan12}. This will let us demonstrate a slightly different analytical approach that will be useful for the readers to master.  The coverage probability analysis for the typical user in both association rules is performed in the following subsections.

\newcommand{\CLi}{\mathbf{Y}}
\subsection{Analysis for Average Power Based Cell Association} \label{sec:HCN_avg}
In this subsection, we derive the coverage probability for a typical mobile user in a $\K$ tier HCN under average power-based cell association \cite{JoSan12}. As was the case in the previous sections, we will first compute the distribution of the distance of the typical mobile user from its serving BS. As discussed above, the closest BS may not always be the serving BS for a given mobile user. It is however easy to see that the serving BS will lie in the set containing closest BS from each tier. Let us denote the serving tier by $S$ and the closest BS of $i\ths$ tier by $B_i$ and its location by $\CLi_i$. Following the discussion in \ref{subsec:NN}, the distribution of the distance $R_i$ of the closest BS $B_i$ of the $i\ths$ tier from the typical mobile user is given by
\begin{align}
f_{R_i}(r)&=2\pi\lambda_i r \exp(-\lambda_i \pi r^2).
\label{eq:f_Ri}
\end{align}
Now given that $B_i$ is located at distance $r$, $B_i$ can be the serving BS only if $B_j$'s $(j\ne i)$ provide lower received power than $B_i$. This leads to the following condition on the distances $R_j$, $j\ne i$,
\begin{align}
p_i R_i^{-\alpha}>p_j R_j^{-\alpha} 
\implies R_j >\left(\frac{p_j}{p_i}\right)^{1/\alpha}r.
\end{align}
Recall that in the single tier case, there was an exclusion region of radius $r$ where no interfering BSs could lie. Similarly, in this case, there is an exclusion region of radius  $e_i(j,r)=\left(\frac{p_j}{p_i}\right)^{1/\alpha}r$
for each tier $j$ in which no interfering BSs of the $j\ths$ tier can lie. Note that $e_i(j,r)$ is defined conditioned on the serving BS being from the $i\ths$ tier.

Now, we will compute the probability $a_i$ that a typical user connects to a BS of the tier $i$. This is known as the {\em association probability} of the $i\ths$ tier. Conditioned on $R_i=r$, the typical user will be connected to $B_i$ if all the other $B_j$'s ($j\ne i$) are located farther away than $e_i(j,r)$:
\begin{align}
\mathbb{P}\left[S=i|R_i=r\right]&= \prob{R_j>\left(\frac{p_j}{p_i}\right)^{1/\alpha}r,\ \forall j\ne i}\notag\\
&\stackrel{(a)}{=} \prod_{j\ne i} \prob{R_j>\left(\frac{p_j}{p_i}\right)^{1/\alpha}r} \stackrel{(b)}{=} \prod_{j\ne i}\exp\left(-\pi\lambda_j\left(\frac{p_j}{p_i}\right)^{2/\alpha}r^{2}\right),
\label{eq:temp23}
\end{align}
where $(a)$ follows from the independence of the $\K$ tiers, and $(b)$ follows from the void probability of a PPP (discussed in detail in Section \ref{subsec:NN}). Therefore, association probability $a_i$ can be computed by averaging \eqref{eq:temp23} over the distribution of $R_i$ (given by \eqref{eq:f_Ri}):
\begin{align}
a_i=\prob{S=i}&=\int_0^\infty \mathbb{P}\left[S=i|R_i=r\right] f_{R_i}(r)\dd r.
\label{eq:a_i}
\end{align}

%
After deriving the distance distributions and the association probabilities, let us derive the joint probability of the event $\{S=i\}$ and the event that the serving BS is located at a distance larger than $r$, which will be useful in the derivation of the coverage probability:
\begin{align}
\prob{R_i>r,S=i}&=\prob{R_i>r, R_j>\left(\frac{p_j}{p_i}\right)^{1/\alpha}R_i,\ \forall j\ne i} \nonumber
\\
&=\int_r^\infty \prob{R_j>\left(\frac{p_j}{p_i}\right)^{1/\alpha}u,\ \forall j\ne i}f_{R_i}(u)\dd u \nonumber\\
&\stackrel{(a)}{=}\int_r^\infty \prod_{j\ne i}\exp\left(-\pi\lambda_j\left(\frac{p_j}{p_i}\right)^{2/\alpha}u^{2}\right) f_{R_i}(u)\dd u \\
&\stackrel{(b)}{=}\int_r^\infty \prod_{j\ne i}\exp\left(-\pi\lambda_j\left(\frac{p_j}{p_i}\right)^{2/\alpha}u^{2}\right)2\pi\lambda_i u \exp(-\lambda_i \pi u^2)\dd u.
\end{align}
where $(a)$ follows on the same lines as \eqref{eq:temp23}, and $(b)$ follows from \eqref{eq:f_Ri}. Using this result, the distribution of the distance of the typical user from its serving BS given that the typical user is connected to $i\ths$ tier is given by
\begin{align}
f_{R_i}(r|S=i)&=\frac{\dd}{\dd r} \prob{R_i>r|S=i}=\frac{\dd}{\dd r} \frac{\prob{R_i>r,S=i}}{\prob{S=i}}\\
&=\frac{1}{a_i}2\pi\lambda_i r \exp(-\lambda_i \pi r^2)\prod_{j\ne i}\exp\left(-\pi\lambda_j\left(\frac{p_j}{p_i}\right)^{2/\alpha}r^{2}\right).
\label{eq:f_R_i_cond}
\end{align}

Using these intermediate results, we are now ready to derive the coverage probability. By definition, the coverage probability in this $\K$-tier HCN case is 
\begin{align}
\pc(\{\tau_i\},\{\lambda_i\},\{\tP_i\})&=\prob{\sinr>\tau_S} \stackrel{(a)}{=} \sum_{i=1}^{\K}\prob{S=i}\prob{\sinr>\tau_i|S=i}\notag \\
&= \sum_{i=1}^{\K}a_i \underbrace{\prob{\sinr>\tau_i|S=i}}_{\mathrm{p}_{i\mathrm{c}}(\Th_i,\{\lambda_i\},\{\tP_i\})} = \sum_{i=1}^{\K} a_i \mathrm{p}_{i\mathrm{c}}(\tau_i,\{\lambda_i\},\{\tP_i\}),
\label{eq:pc_interm1}
\end{align}
where $(a)$ follows from the total probability law. Since we already derived an expression for $a_i$ in \eqref{eq:a_i}, we just need to compute per-tier coverage probability $\mathrm{p}_{i\mathrm{c}}(\tau_i,\{\lambda_i\},\{\tP_i\})$ in order to complete our derivation. We do this next:  
\begin{align}
\mathrm{p}_{i\mathrm{c}}(\Th_i,\{\lambda_i\},\{\tP_i\})&= \prob{\sinr>\tau_i|S=i} = \E_{R_i}\left[ \vphantom{\frac33} \PP[\sinr >\Th_i\ |\  R_i=r,S=i] \right]\nonumber\\
& = \int_{r>0} \PP[\sinr > \Th_i\ |\  R_i=r,S=i ]f_{R_i}(r|S=i)\dd r\label{eq:temp27}. 
\end{align}
where the unconditioning in the last step with respect to distance $R_i$ needs to be done using conditional distribution $f_{R_i}(r|S=i)$ given by \eqref{eq:f_R_i_cond}. Now substituting the above expression in \eqref{eq:pc_interm1}, we get
\begin{align}
&\pc(\{\Th_i\},\{\lambda_i\},\{\tP_i\})  =\sum_{i=1}^{\K} a_i \int_{r>0} \PP\left[\frac{\rh_{\CLi_i}\tP_i R_i^{-\alpha}}{\sigma^2+I} > \Th_i\ \Big| \ R_i=  r, S=i\right]  \frac{2\pi\lambda_i}{a_i} r e^{-\pi\sum_{j}\lambda_j\left(\frac{p_j}{p_i}\right)^{2/\alpha}r^{2}} \dd r \nonumber \\
& =  \sum_{i=1}^{\K}   2\pi \lambda_i\int_{r>0} re^{-\pi\sum_{j}\lambda_j\left(\frac{p_j}{p_i}\right)^{2/\alpha}r^{2}} \PP[\rh_{\CLi_i}>\Th_i\tP_i^{-1}r^{\alpha}(\sigma^2+I) \ |\ S=i,R_i=r ]  r \dd r. \label{eq:HetNet1pc}
\end{align}
Using the fact that $\rh_{\CLi_i} \sim \exp(1)$,   the inner probability term can be  further simplified as
\begin{align}
\PP[\rh_{\CLi_i}>\Th_i\tP_i^{-1}r^{\alpha}(\sigma^2+I)\ |\ R_i= r,S=i] &=   e^{-\tP_i^{-1} \Th_i r^{\alpha} \sigma^2}{\cal L}_{I}(\Th_i\tP_i^{-1}r^{\alpha} | R_i= r,S=i)\label{eq:HetNet1LC},
\end{align}
where ${\cal L}_{I}(s| R_i= r,S=i)$ is the conditional interference Laplace transform which is the last component that needs to be computed. For notational simplicity, we will denote this conditional Laplace transform by ${\cal L}_{I}(s)$ with the understanding that this is conditioned on the event $\{R_i= r,S=i\}$. 
Note that interference power experienced by the typical mobile user in the HCN case, as evident from \eqref{eq:SINR_HCN}, is summation of the interference power $I_j$ from each tier. Hence, the Laplace transform of the interference is given as
 \begin{align}
{\cal L}_{I}(s)&=\E\left[\exp(-I(s))\right]=\E\left[\exp(-\sum_{j=1}^\K I_j(s))\right]\nonumber\\
&=\prod_{j=1}^\K\E\left[\exp(-I_j(s)\right]=\prod_{j=1}^\K{\cal L}_{I_j}(s) \label{eq:HetNet1Laplaceproduct}
\end{align}
where ${\cal L}_{I_j}(s)$ is the Laplace transform of the interference power from the BSs belonging to the $j\ths$ tier. Note that the second to the last step is due to the independence assumption among tiers.
Now, similar to the interference in the single tier case, the interference $I_j$ from the $j\ths$ tier is also a standard $M/M$ shot noise created by a PPP of intensity $\lambda$ outside a disc of radius $e_i(j,r)$ centered at the origin $\mathbf{o}$. As discussed earlier in this section, the exclusion radius $e_i(j,r)$ is different across tiers. Using this exclusion radius, we can derive the Laplace transform of the $j\ths$ tier interference on the same lines as Section \ref{sec:Int_singletier}. The final expression is the same as \eqref{eq:laplaceexp} with exclusion radius being $e_i(j,r)$ and is given below:
\begin{align}
{\cal L}_{I_j}(s)&=  \exp\left(-2\pi\lambda_j \int_{e_i(j,r)}^{\infty} \left(\frac{1}{1+(s\tP_j)^{-1} x^{\alpha}}\right) x  \dd x \right).
\label{eq:laplaceHetNet1}
\end{align}
Now substituting this result in \eqref{eq:HetNet1Laplaceproduct}, ${\cal L}_{I}(\Th_i\tP_i^{-1}r^{\alpha})$ can be computed as
\begin{align}
{\cal L}_{I}(\Th_i\tP_i^{-1}r^{\alpha})&=\prod_{j=1}^\K \exp\left(-2\pi\lambda_j \int_{e_i(j,r)}^{\infty} \left(\frac{1}{1+(\Th_i\tP_j/\tP_i)^{-1} x^{\alpha}/r^{-\alpha}}\right) x  \dd x \right).
\end{align}
Employing change of variables $u = \left(x/r\right)^2(\Th_i\tP_j/\tP_i)^{-\frac{2}{\alpha}}$ results in
\begin{align}
{\cal L}_{I}(\Th_i\tP_i^{-1}r^{\alpha})&=\prod_{j=1}^\K \exp\left(-\pi\lambda_j
r^{2} (\Th_i\tP_j/\tP_i)^{\frac{2}{\alpha}}
  \int_{ \Th_i^{-\frac{2}{\alpha}}}^{\infty} \left(\frac{1}{1+u^{\alpha/2}}\right) u  \dd u \right).
\end{align}
Now using the definition of function $\rho(\cdot,\cdot)$ from \eqref{eq:rho}, we get
\begin{equation}
{\cal L}_{I}(\Th_i\tP_i^{-1}r^{\alpha}) = \exp\left(-\pi \lambda_j \sum_{j=1}^\K r^{2} (\tP_j/\tP_i)^{\frac{2}{\alpha}}  \rho(\Th_i,\alpha) \right)
\label{eq:Ir122}.
\end{equation}
Combining \eqref{eq:HetNet1pc}, \eqref{eq:HetNet1LC} and \eqref{eq:Ir122}, we get the following Theorem.

\begin{theorem}\label{thm:hetnetavgpower}
The downlink coverage probability for a typical mobile user in an open access k-tier HCN with average power based cell selection is
\begin{align}
&\pc(\{\Th_i\},\{\lambda_i\},\{\tP_i\}) = \nonumber \\
& \sum_{i=1}^{\K}   2\pi \lambda_i\int_{r>0} r
\exp\left(-\tP_i^{-1} \Th_i r^{\alpha} \sigma^2\right)
\exp\left(-\pi \sum_{j=1}^\K \lambda_j r^{2} (\tP_j/\tP_i)^{\frac{2}{\alpha}} (1+\rho(\Th_i,\alpha)) \right)  r \dd r.
\end{align}
\end{theorem}

\subsubsection{Special Cases}
As in single tier case studied in Section~\ref{sec:DL}, we now consider some special cases  for HCNs which offer further simplifications of Theorem \ref{thm:hetnetavgpower} and additional insights.

\subsubsubsection{(i) Interference-limited (No-noise) Case}
The general downlink coverage result of Theorem \ref{thm:hetnetavgpower} can be specialized to a practically relevant case of interference-limited networks ($\sigma^2 = 0$) as follows.

\begin{cor}
In an interference-limited network ($\N =0$), the downlink coverage probability of a typical mobile user in a $\K$-tier HCN with average power-based cell association simplifies to
\begin{align}
\pc(\{\lambda_i\}, \{\T_i\}, \{\tP_i\}) =\frac{\sum_{i=1}^\K\lambda_i(\tP_i)^{\frac{2}{\alpha}}}{\sum_{j=1}^\K \lambda_j  (\tP_j)^{\frac{2}{\alpha}} (1+\rho(\Th_i,\alpha)) }\label{eq:hetnetavgpowersimplecaseI}.
\end{align}
\end{cor}

Comparing \eqref{eq:hetnetavgpowersimplecaseI} and \eqref{eq:SIR-1tier}, the results are identical, the difference being  additional terms due to multiple tiers with different parameters.

\subsubsubsection{(ii) Same per-tier SIR threshold}
If the per tier SIR thresholds are the same  ($\T_i = \T ~ \forall ~ i$), which is quite reasonable, it can be observed that the coverage probability simplifies to:
\begin{equation}
\pc(\{\lambda_i\}, \T, \{\tP_i\}) = \frac{1}{ 1+\rho(\Th,\alpha)}.
\label{eq:hetnetavgpowersimplecaseII}
\end{equation}

 This result indicates that the outage probability is now independent of the BS transmit power $\{\tP_i\}$ and BS density $\{\lambda_i\}$. It can be intuitively seen by the fact that as we change the density of BSs by some factor, it impacts the distances to the serving and all interfering BSs by the same factor, leaving the SIR invariant to the changes in density. Furthermore, not surprisingly, with $\K=1$, i.e. the single tier case, we also get the same result as obtained in \eqref{eq:SIR-1tier}.


\subsection{Analysis for Instantaneous Power Based Cell Selection} \label{sec:HCN_inst}
After discussing the average power-based cell selection in the previous subsection, we now consider instantaneous power-based cell selection in this subsection.
Under this cell association rule, the typical user at the origin is in coverage if 
\[  \max_{\rX \in \Phi_i}\{\sinr(\rX)\} > \T_i, \]
for some $1\leq i\leq \K$, where $\sinr(\x)$ is given by~\eqref{eq:SINR_HCN}. The key challenge in the downlink analysis under this cell association scheme is that fact that the closest BS from each tier is not necessarily the serving BS due to the presence of fading. This does not allow us to {\em fix} the serving BS in the same way as we did in the previous subsection. Recall that we assume open access network where the typical mobile mobile user is allowed to connect to any BS in the network. Under this assumption, the mobile user is said to be in coverage if the received $\sinr$ from at least one BS is higher than its corresponding target $\Th_i$. This can be mathematically expressed as: 
\begin{align}
\pc(\{\lambda_i\}, \{\T_i\}, \{\tP_i\}) 
&= \P \left(\bigcup_{i\in \{1,2,\ldots \K\}, \rX_i \in \Phi_i} \sinr(\rX_i) > \T_i \right) \notag \\
&=\E\left[ \i \left( \bigcup_{i\in \{1,2,\ldots \K\}, \rX_i \in \Phi_i} \sinr(\rX_i) > \T_i  \right) \right] .
\end{align}
As demonstrated first in~\cite{DhiGan12}, the analysis greatly simplifies for $\T_i>1$ (0 dB) for all tiers. This is because under this assumption at most one BS across all $\K$ tiers can satisfy the coverage condition, i.e., at most one BS can provide $\sinr$ greater than the required $\sinr$ threshold. To understand this intuitively, consider a system with two BSs. Denote the received powers from these two BSs at the mobile user by $a_1$ and $a_2$. Depending upon the choice of the serving BS, the received $\sinr$ will be either $\frac{a_1}{a_2 + \sigma^2}$ or $\frac{a_2}{a_1 + \sigma^2}$. If $\frac{a_1}{a_2 + \sigma^2} > 1$, it means $a_1 > a_2 + \sigma^2$, which in turn implies $a_1 > a_2$. Therefore, $\frac{a_2}{a_1} < 1$, which further implies that the received $\sinr$ from the other BS is $\frac{a_2}{a_1 + \sigma^2} < 1$. In short, only one of the two $\sinr$s can be larger than $1$ (0 dB). Thus if we enforce the target $\sinr$s for all the tiers to be larger than $1$, the coverage condition can be met by at most one BS. As an aside, note that this result can actually be extended to show that at most $m$ BSs can meet the coverage condition if the target $\sinr$ is greater than $1/m$ for any positive integer $m$. Interested readers are referred to \cite[Lemma 1]{DhiGan12} for more details. 

For the rest of this Section, we assume $\T_i>1$, $\forall\ i$, which means the typical mobile user can connect to at most one BS (as discussed above). Under this condition, the coverage probability can be expressed in terms of the sum of the indicator functions as follows:
\begin{align}
\pc(\{\lambda_i\}, \{\T_i\}, \{\tP_i\}) &= \E\left[ \i \left( \bigcup_{i\in \{1,2,...\K\}, \rX_i \in \Phi_i} \sinr(\rX_i) > \T_i  \right) \right]\nonumber\\
&\stackrel{(a)}{=}\sum_{i=1}^{\K}\E \left[\sum_{\rX_i \in \Phi_i}  \left[\i\left( \sinr(\rX_i) > \T_i  \right)  \right]\right],\label{eq:pcdef}
\end{align}
where $(a)$ is in general an upper bound (by {\em union bound}) but holds with equality if at most one of the BSs satisfy the coverage condition, which is precisely the case when we assume $\T_i>1$, $\forall\ i$. We first compute the inner expectation, which can be written as
\begin{align}
\mathrm{p}_{\mathrm{c}i}=\E \left[\sum_{\rX_i \in \Phi_i}  \left[\i\left( \sinr(\rX_i) > \T_i  \right)  \right]\right]=\E \left[\sum_{\rX_i \in \Phi}  \left[\i\left( \sinr(\rX_i) > \T_i  \right)  \i\left(\rX_i\in\Phi_i\right)\right]\right],\label{eq:pckdef}
\end{align}
where we defined $\Phi=\cup_{i=1}^\K \Phi_i$ for notational simplicity. The reason for converting the summation over $\Phi_i$ to $\Phi$ will be clear shortly. Note that the above expression is simply the expected value of the sum of the function $f(\rX_i,\Phi\setminus \{\rX_i\})=\i\left( \sinr(\rX_i) > \T_i  \right)\i\left(\rX_i\in\Phi_i\right)$ over all points of the PPP $\Phi$. Since function $f(\cdot)$ is dependent on both $\rX_i$ and point process $\Phi\setminus \{\rX_i\}$, we cannot apply the standard Campbell's theorem that we discussed in Section~\ref{sec:background}. Note that the dependence of function $f(\cdot)$ on $\Phi\setminus \{\rX_i\}$ is because the interference $I(\Phi \setminus \{\rX_i\})$ observed at the typical user is the function of the received powers from all the BSs, except the serving BS located at $\rX_i$. In order to evaluate the above expression we need to use the closely related Campbell-Mecke theorem, which for the homogeneous PPP can be expressed as~\cite[Equation (4.71)]{ChiStoB2013}: 
\begin{align}
\mathbb{E}\left[\sum_{\rX_i\in \Phi} f(\rX_i,\Phi \setminus \{\rX_i\} )\right]&=\lambda\int_{\mathbb{R}^2} \mathbb{E}\left[f(\x_i,\Phi)\right] {\rm d}  \x_i.
\label{eq:CM}
\end{align}
Let us first compute $\mathbb{E}\left[f(\x_i,\Phi)\right]$ below:
\begin{align}
\mathbb{E}\left[f(\x_i,\Phi)\right] &= \mathbb{E} \left[ \i\left(\frac{\tP_i \rh_{\x_i} \|\x_i\|^{-\alpha}}{I(\Phi)+ \sigma^2} > \T_i  \right)\i\left(\x_i\in\Phi_i\right) \right]\\
&= \mathbb{P}\left(\frac{\tP_i \rh_{\x_i} \|\x_i\|^{-\alpha}}{I(\Phi)+ \sigma^2} > \T_i  \right)\mathbb{P}\left(\x_i\in\Phi_i\right)\\
&= \frac{\lambda_i}{\lambda} \mathbb{P}\left(\frac{\tP_i \rh_{\x_i} \|\x_i\|^{-\alpha}}{I(\Phi)+ \sigma^2} > \T_i  \right),
\label{eq:E_f}
\end{align}
where note that the ``interference'' term $I (\Phi)$ is now sum of received powers from {all} points of the PPP (i.e., point $x_i$ is not excluded), which follows from the Campbell-Mecke theorem given by \eqref{eq:CM}. Also, the $\frac{\lambda_i}{\lambda}$ factor is due to the  term $\mathbb{P}\left(\x_i\in\Phi_i\right)$ which is equal to  $\frac{\lambda_i}{\lambda}$ from the independent thinning theorem. Now combining \eqref{eq:pckdef}, \eqref{eq:CM}, and \eqref{eq:E_f}, we get
\begin{align}
\mathrm{p}_{\mathrm{c}i}
&= \lambda\int_{\R^2} \frac{\lambda_i}{\lambda} \P\left(\frac{\tP_i \rh_{\x_i} \|\x_i\|^{-\alpha}}{I(\Phi)+\N} > \T_i  \right) \d \x_i =\lambda_i\int_{\R^2}\P\left(\frac{\tP_i \rh_{\x_i} \|\x_i\|^{-\alpha}}{I(\Phi)+\N} > \T_i  \right) \d \x_i.
\label{eq:pcval}
\end{align}
Substituting \eqref{eq:pcval} into \eqref{eq:pckdef}, we get
\begin{align}
\pc(\{\lambda_i\}, \{\T_i\}, \{\tP_i\}) 
&=\sum_{i=1}^{\K} \lambda_i\int_{\R^2}\P\left(\frac{\tP_i \rh_{\x_i} \|\x_i\|^{-\alpha}}{I(\Phi)+\N} > \T_i  \right) \d \x_i.
\end{align}
For notational simplicity, we denote  $I(\Phi) = \sum_{j=1}^{\K} \sum_{\rX_j \in \Phi_j} \tP_j \rh_{\rX_j} \|\rX_j\|^{-\alpha}$ by simply $I$. Now using the fact that $\rh_{\rX_i} \sim \exp(1)$, we can express coverage probability in terms of the Laplace transform of $I$ as follows: 
\begin{align}
\pc(\{\lambda_i\}, \{\T_i\}, \{\tP_i\}) 
&= \sum_{i=1}^{\K} \lambda_i \int_{\R^2} \L_{I} \left(\frac{\T_i}{\tP_i \|\x_i\|^{-\alpha}}  \right) \exp{\left(\frac{-\T_i \N}{\tP_i \|\x_i\|^{-\alpha}}\right)} \d \x_i
\label{eqn:Pc_der}
\end{align}
The Laplace transform of interference can now be computed as follows: 
\begin{align}
\L_{I}\left(s\right)
&= \E_{I} \left[\exp \left(-s I\right)   \right] =  \E \left[ \exp \left(-s \sum_{j=1}^{\K} \sum_{\rX_j \in \Phi_j} \tP_j \rh_{\rX_j} \|\rX_j\|^{-\alpha}   \right)  \right]\\
&= \E \left[ \prod_{j=1}^{\K} \prod_{\rX_j \in \Phi_j} \exp \left(-s \tP_j \rh_{\rX_j} \|\rX_j\|^{-\alpha}   \right) \right] \stackrel{(a)}{=} \prod_{j=1}^{\K} \E \left[  \prod_{\rX_j \in \Phi_j} \exp \left(-s \tP_j \rh_{\rX_j} \|\rX_j\|^{-\alpha}   \right) \right] \\
&\stackrel{(b)}{=} \prod_{j=1}^{\K} \E_{\Phi_j} \left[  \prod_{\rX_j \in \Phi_j} \E_h \left[ \exp \left(-s \tP_j \rh \|\rX_j\|^{-\alpha}   \right) \right] \right] \stackrel{(c)}{=} \prod_{j=1}^\K \E_{\Phi_j}\left[\prod_{\rX_j\in\Phi_j} \frac{1}{1+s\tP_j \|\rX_j\|^{-\alpha}}  \right],
\end{align}
where $(a)$ follows from the independence of PPPs modeling different tiers, $(b)$ follows from the independence of the fading variables $\rh_{\rX_j}$'s (generic fading variable is denoted by $\rh$ in this step), and $(c)$ follows from the Rayleigh fading assumption, i.e., $\rh \sim \exp(1)$. Now we can use the PGFL of PPP to convert product over the PPP to an integral as follows
\begin{align}
\L_{I}\left(s\right)&=  \prod_{j=1}^\K \exp  \left( -\lambda_i \int_{\R^2} \left(1-\frac{1}{1+s\tP_j \|\x_j\|^{-\alpha}}   \right) \d \x_j  \right).  
\end{align}
Converting from cartesian to polar coordinates $\x_j=(r,\theta)$, we get
\begin{align}
\L_{I}\left(s\right)
&= \prod_{j=1}^\K \exp  \left( -2 \pi \lambda_i \int_{0}^{\infty} \left(1-\frac{1}{1+s\tP_j r^{-\alpha}}   \right) r \d r  \right). 
\end{align}
To simplify this integral further, we perform the change of variable $(s\tP_j)^{-\frac{1}{\alpha}}r\rightarrow u$, which gives 
\begin{align}
\L_{I}\left(s\right)&= \prod_{j=1}^\K \exp  \left( -2 \pi \lambda_i (s \tP_j)^{2/\alpha} \int_{0}^{\infty} \left(1-\frac{1}{1+u^{-\alpha}}   \right) u\d u  \right) \nonumber \\
&= \prod_{j=1}^\K \exp  \left( -2 \pi \lambda_i (s \tP_j)^{2/\alpha} \int_{0}^{\infty} \left(\frac{1}{u^{\alpha} + 1}   \right) u \d u \right) \nonumber \\
&=\exp\left(-s^{2/\alpha}\Cfunc(\alpha)\sum_{i=1 }^\K \lambda_i \tP_i^{2/\alpha}\right),
\label{eqn:L}
\end{align}
 where $\Cfunc(\alpha) = \frac{2\pi^2}{\alpha}\csc(\frac{2\pi}{\alpha})$. Substituting this expression in \eqref{eqn:Pc_der} gives the final expression for coverage probability $\pc(\{\lambda_i\}, \{\T_i\}, \{\tP_i\})$, which is given next.

\begin{theorem} For $\T_i>1$, the downlink coverage probability for a typical mobile user in an open access $\K$-tier HCN under maximum instantaneous power-based cell selection is 
\begin{equation}
\pc(\{\lambda_i\}, \{\T_i\}, \{\tP_i\})= 2\pi\sum_{i=1}^{\K} \lambda_i \int_{0}^{\infty}\exp{\left(-x^2\left(\frac{\T_i}{\tP_i}\right)^{2/\alpha}\Cfunc(\alpha)\sum_{m=1 }^\K \lambda_m \tP_m^{2/\alpha}  \right)}\exp{\left(-\frac{\T_i \N}{\tP_i }x^\alpha \right)}  x\d x,
\label{eqn:Pc}
\end{equation}
where $\Cfunc(\alpha) = \frac{2\pi^2}{\alpha}\csc(\frac{2\pi}{\alpha})$.
\label{Thm:1}
\end{theorem}


\subsubsection{Special Cases}
As in the macrocell-only case studied in Section~\ref{sec:DL} and the HCN under average-power cell selection rule studied in Section~\ref{sec:HCN_avg}, we now consider some special cases  for the instantaneous power based cell selection which offer further simplifications of Theorem \ref{Thm:1} and additional insights.

\subsubsubsection{(i) Interference-limited (No-noise) Case}
The general downlink coverage result of Theorem \ref{Thm:1} can be specialized for interference-limited networks ($\sigma^2 = 0$) as follows. This case is particularly relevant for HCNs, which are often interference-limited due to dense and organic deployment of small cells.
\begin{cor}
\label{Cor:1}
In an interference-limited network,  \ie, $\N =0$, the downlink coverage probability of a typical mobile user simplifies to
\[\pc(\{\lambda_i\}, \{\T_i\}, \{\tP_i\}) = \frac{\pi}{\Cfunc(\alpha)}  \frac{\sum_{i=1}^\K \lambda_i \tP_i^{2/\alpha} \T_i^{-2/\alpha}}{\sum_{i=1}^\K \lambda_i \tP_i^{2/\alpha}   }, \ \ \T_i>1.\]
\end{cor}

This follows immediately from Theorem \ref{Thm:1} with $\sigma^2=0$.  This is a remarkably simple closed-form expression, which is useful in understanding how coverage probability depends upon various system parameters, such as the transmit powers of different tiers.

\subsubsubsection{(ii) Same per-tier SIR thresholds}
If the per tier SIR thresholds are the same  ($\T_i = \T ~ \forall ~ i$), which is quite reasonable, it can be observed that the coverage probability simplifies to:
\begin{equation}
\pc(\lambda, \T, \tP) = \frac{\pi}{\Cfunc(\alpha)\T^{2/\alpha}}.
\label{eq:SIR-simple}
\end{equation}
%
From the above result, we can observe that the coverage probability is independent of the density of the BSs, number of tiers, their respective powers which indicates that the SIR distribution is invariant of density and transmit powers.  The same result was also obtained for the average power based association case in the previous subsection. 

Comparing \eqref{eq:SIR-simple} with \eqref{eq:hetnetavgpowersimplecaseII}, they are  related, but not identical due to the different association rules considered.  Namely, here we allow association to the highest \emph{instantaneous SINR}, including fading, and assume that the threshold for success is $\T_i > 1$. We will comment on the extensions to $\T_i < 1$ in the next Section. Please note that we chose to look at the instantaneous power-based association rule to demonstrate additional tools (such as the Campbell-Mecke theorem) that did not explicitly appear in the previous two sections. Finally, please note that while the discussion on HCN was limited to the downlink case, the analysis of uplink does not require any new additional tools (besides the ones introduced in this paper). Interested readers are advised to refer to \cite{SinZhaJ2015,LeeSanJ2014,DiGuaJ2016,MarGomJ2016} for more details.


\section{Extensions}\label{sec:ext}
The model and results in this tutorial have been extended and generalized in many directions.  Without any hope of discussing them all, we focus now on a few important classes of such generalization.   Please refer to \cite{ElSawyHossainHaenggi2013,MukBook} for a more comprehensive summary of the recent results, especially on HCNs. 


\subsection{Incorporating general channel models} \label{sec:GenChan}

The assumption of Rayleigh fading (exponential distribution for the channel power gain) leads to a particularly convenient form for the coverage probability in terms of the Laplace transform of interference power distribution, which is in general easier to characterize than the probability density function of interference power~\cite{SouSilJ1990,HaenggiBook}. Technically, assuming Rayleigh fading for only  the {\em serving link} is sufficient to express the coverage probability in terms of the Laplace transform of interference, which means the fading distribution for the interfering links can be generalized without loss of tractability, as done in \cite{AndBac11}. The generalization of the fading distribution for the serving link is usually much more complex~\cite{Brown2014,MadResJ2016}. One key exception is when the channel power gain of the serving link is chosen from the {\em exponential family}, in particular, Gamma distribution (e.g., Nakagami-$m$). In the case of Nakagami-$m$, the coverage probability can be expressed in terms of the higher-order derivative of the Laplace transform of interference, which is relatively easier to evaluate compared to a more general distribution~\cite{TorValJ2012,TanDhiJ2014}. 

In addition to the direct extension discussed above, note that the effect of general fading and shadowing distributions can also be incorporated by treating them as {\em equivalent} random perturbations in the locations of the BSs. Using the displacement theorem, their effect on the received power can be incorporated by transforming the original  homogeneous PPP of the BSs into a new homogeneous PPP whose density depends upon the fractional moment of the shadowing distribution. This was discussed in Section~\ref{sec:shadowing} in the context of incorporating shadowing in the downlink analysis. For the application of this general idea to other scenarios, please refer to~\cite{Blaszczyszyn2013,DhiAndJ2014,Brown2014,MadResJ2016,BlaKeeC2013,KeeBlaC2013,ZhaHaeJ2014} and the references therein.

Finally, while power law pathloss model with an arbitrary exponent $\alpha>2$ is a good starting point and quite common in both theory and practice, more general pathloss models have also been incorporated in the analyses. One way is again to transform the original PPP to an equivalent PPP where the effect of more general pathloss model will appear in the density of the transformed PPP~\cite{Brown2014}.   Another is to use a multislope power law path loss model where the path loss exponent increases as a function of the link distance \cite{ZhaAnd15,DinWanJ2016}.  The multislope model is more flexible and accurate, and can have interesting implications particularly for dense networks \cite{AndDur16}.

\subsection{Various cell selection strategies}

There are three general classes of cell selection strategies: (i) {\em average power based}, where the receivers connect to the BSs providing maximum average received power (considered for downlink analysis of macro-only case in Section~\ref{sec:DL} and $\K$-tier HCN case in Section~\ref{sec:HCN_avg}), (ii) {\em instantaneous power based}, where the receivers connect to the BSs providing maximum instantaneous received power (considered for the $\K$-tier HCN analysis in Section~\ref{sec:HCN_inst}), and (iii) {\em strategies with cell selection bias}, where an additional parameter, termed the {\em cell selection bias}, is considered along with the received power, e.g., see \cite{SinDhi13}. While we discussed the first two cell selection strategies for HCN, including cell selection bias does not require any new tools. Interested readers can refer to \cite{JoSan12, SinDhi13} for more details. We also only considered open-access case for HCNs in this tutorial, where users are allowed to connect to any BS in the network. This requirement of open-access is often not satisfied, in particular for femtocells and if WiFi is introduced in a so-called multi-RAT (radio access technology) network. Reference \cite{DhiGan12} also handles closed-access (which reduces the coverage probability and violates the SIR invariance property) and the multi-RAT case is considered in \cite{SinDhi13}.

The rationale for cell selection bias stems from the desire to maximize the per link data rate, which is related to but is not necessarily equivalent to the SINR.  Particularly in an HCN, too many of the mobiles connect to a macrocell, because of its high power, leaving the small cells underutilized and the macrocells heavily congested.  The selection bias is a simple and crude but remarkably effective tool for pushing mobiles onto the small cells, and has been observed to achieve near-optimum results when the bias is chosen correctly \cite{SinDhi13,YeRon13}.

These three strategies may appear to the different at the surface, but are fundamentally related. For instance, it can be shown that the maximum instantaneous power based cell selection in the presence of general fading is equivalent to the maximum power based cell selection in the absence of fading in an equivalent PPP (where the effect of fading has been incorporated in the density of the equivalent PPP)~\cite{Brown2014}. This general idea was introduced in the context of including the effect of shadowing in the performance analysis of downlink cellular networks in Section~\ref{sec:shadowing}.

We conclude this discussion about cell selection strategies by revisiting the assumption in Section~\ref{sec:HCN} that $\T_i> 1$ (0 dB) for all tiers. This assumption allowed us to use union bound in the coverage analysis with equality, thereby lending tractability to the downlink HCN analysis. The analysis has been extended to  $\T_i < 1$ in \cite{DhiGanC2011a,KeeBlaC2013} using two entirely different approaches. Naturally, the resulting expressions in both the cases are not as nice as the ones that we get under the assumption $\T_i> 1$, $\forall i$. The extension in \cite{DhiGanC2011a} is based on deriving the joint distribution of interference and maximum signal strength using tools from \cite{NguBacC2010}. On the other hand, the extension in \cite{KeeBlaC2013} is based on deriving the $k$-coverage result for cellular networks.

\subsection{More general spatial setups}

BSs are not independently placed in practice, but the results given here can in principle be generalized to point processes that model repulsion or minimum distance, such as determinantal \cite{LiBacAnd15} and Mat{\'e}rn processes \cite{ElSHosJ2014a}.  Other recent results from Haenggi indicate that a fixed positive SINR shift (e.g. about 2 or 3 dB) can be applied to the results obtained under the PPP assumption to yield accurate results for other more general point processes, or even the hexagonal grid \cite{GuoHae15}. 

On the same lines as above, the BS locations across tiers in an HCN also exhibit dependence. For instance, an operator will likely not deploy a picocell very close to a macrocell. This repulsion has been modeled in~\cite{DenZhoJ2015} using a Poisson Hole Process (PHP)~\cite{YazDhiJ2016,LeeHaeJ2012}. The idea is to carve out holes around the macro BS locations in which certain types of small cells, such as picocells, cannot be deployed. The small cell locations in such a setup are modeled by a PHP where the holes are driven by the macrocell locations. 

Finally, throughout this tutorial, we assumed that the user locations are independent of the BS locations. While this is a preferred case for the analysis of coverage-centric deployments, as was the case in conventional macro-only networks, this is not quite accurate for the analysis of capacity-centric deployments, where the BSs are deployed at the areas of high user density (and hence high data traffic). In such cases, it is important to model correlation in the BS and user locations. This correlation has recently been modeled in \cite{SahDhiJ2016} using a Poisson cluster process (PCP) model where the small cells are assumed to lie at the cluster center around which the mobile users form clusters with general distribution. The analysis is enabled by new distance distributions and analytical tools recently presented in \cite{AfsDhiJ2016,AfsDhiJ2016a}. On similar lines, if small cell deployments exhibit clustering,  they can also be modeled using PCP, as done in \cite{ChuHasJ2015}.

\subsection{Multiple-input multiple-output HCNs}

While this tutorial assumed single-antenna per wireless node (both BSs and users), the analysis has already been extended in several important ways in the literature to incorporate multiple antennas per BSs and/or users. One fundamental consideration in networks with multi-antenna nodes is that the received power distribution from a given wireless node (BS or a user) will be different depending upon whether that node is a serving node or an interfering node (due to the so-called beamforming or the precoding gain). This consideration alone complicates the analysis of MIMO HCNs significantly. For instance, as discussed in detail in \cite{DhiKou13} for the downlink MIMO HCNs, if a BS equipped with $m$ antennas serves a single user per resource block, it can be shown under Rayleigh fading and zero-forcing precoding that the fading component of the effective channel power gain for the serving link is Gamma distributed with shape parameter $m$ and scale parameter $1$. The fading component of the channel power gain from similar interfering BSs can be shown to be exponentially distributed. Therefore, the effect of MIMO transmission techniques can be effectively captured by considering different effective fading distributions for the serving and interfering links. 

As discussed above, under Rayleigh fading, the channel power distribution of the serving link can be modeled as Gamma, which as discussed in Section~\ref{sec:GenChan} will reduce the coverage probability to a higher-order derivative of the Laplace transform of interference. While this derivative can be evaluated using Faa di Bruno's rule along with Bell polynomials, as done in~\cite{DhiKou13,GupDhiVisAnd14}, exploiting the recursive structure of the expression, the coverage probability can also be expressed as a function of lower triangular Toeplitz matrix as presented in \cite{LiZhaJ2014}. Similar approach has been followed in the performance analysis of multi-tier cellular system in \cite{LiZhaJ2016}. 


\subsection{Millimeter Wave (mmWave) communication systems}

The stochastic geometry tools introduced in this tutorial can also be used to analyze wireless systems operating at mmWave frequencies (above 30 GHz) \cite{Bai2014,Singh2015}.  These high frequency systems exhibit two fundamental differences compared to sub 6 GHz systems. The first difference is that the {\em blocking} of wireless signals as a result of various obstacles, such as buildings and foliage, is much more severe in mmWave systems. This necessitates the separate modeling of Line-of-Sight (LOS) and non-LOS (NLOS) links. For tractability, it is often assumed that the wireless link to a given BS is either LOS or NLOS with a certain probability independent of the state of the links to other BSs. The probability with which a BS is NLOS (also termed {\em blocking probability}) is dependent on the distance between the BS and the receiver of interest. The effect of blocking can be modeled using random shape theory \cite{BaiVaze2014} or a LOS ball model \cite{Bai2014}.  Ultimately, given a typical user, the BS locations can be modeled as a superposition of two non-homogeneous PPPs, each modeling the LOS and NLOS BSs. 

The second key difference is that mmWave systems will have large antenna arrays (each element being very small) which result in highly directional communication. Hence, Nakagami-$m$ fading with a large degree of freedom is more suitable for modeling mmWave channels compared to Rayleigh fading \cite{Bai2014}. The non-homogeneity of the BS PPP process (consequence of blocking) and general Nakagami-$m$ fading adds significant complexity to the analyses.
The analysis can be extended to multi-tier mmWave systems \cite{Renzo2015} and the related multi-operator mmWave systems \cite{GupAndHea16} through the superposition of more than one BS PPP.  Given the level of interest and activity in mmWave cellular systems, we expect the application of stochastic geometry to mmWave communication to be a vibrant area of research in the coming years.



\section{Acknowledgments}

The authors gratefully acknowledge the indispensable contributions of F. Baccelli to these results.  We also recognize our co-authors R. K. Ganti, T. Novlan, S. Singh, and X. Zhang for their contributions to the original works.  We also thank M. Afshang and P. Parida for their inputs and proofreading of the manuscript.

\bibliographystyle{IEEEtran}
\bibliography{CellularPPP_tutorial_v24.bbl}
\newpage


\end{document}